\documentclass[aps,physrev,twocolumn, superscriptaddress]{revtex4-2}

\usepackage{graphicx}
\usepackage{hyperref}
\usepackage{siunitx}
\usepackage{amssymb}
\usepackage{amsmath}
\usepackage{cleveref} 
\usepackage{pifont}
\usepackage{acro}
\usepackage{mathrsfs}
\usepackage{pifont}
\usepackage{xcolor}

\newcommand{\cmark}{\ding{51}}%
\newcommand{\xmark}{\ding{55}}%

\DeclareAcronym{scr}{short=SCR, long=strong coupling regime}
\DeclareAcronym{cpf}{short=CPF, long=controlled phase flip}
\DeclareAcronym{cqed}{short=CQED, long=cavity quantum electrodynamics}
\DeclareAcronym{dit}{short=DIT, long=dipole induced transparency}
\DeclareAcronym{ppm}{short=ppm, long=parts per million}
\DeclareAcronym{fwhm}{short=FWHM, long=full-width at half-maximum}
\DeclareAcronym{dof}{short=DoF, long=degrees of freedom}
\DeclareAcronym{pdl}{short=PDL, long=polarization dependent loss}
\DeclareAcronym{pmd}{short=PMD, long=polarization mode dispersion}
\DeclareAcronym{re}{short=RE, long=remote entanglement}
\DeclareAcronym{jch}{short=JCH, long=Jaynes-Cummings interaction Hamiltonian}
\DeclareAcronym{npbs}{short=NPBS, long=non-polarizing beamsplitter}
\DeclareAcronym{na}{short=NA, long=numerical aperture}

\begin{document}


\title{Design Tradeoffs in Photonically Linked Qubit Networks}


\author{Ely Novakoski}
\email[]{Contact Author: Ely.Novakoski@Duke.edu}
\affiliation{Duke Quantum Center, Duke University, Durham, NC, USA}
\affiliation{Department of Electrical and Computer Engineering, Duke University, Durham, NC, USA}
\author{Jungsang Kim}%
\affiliation{Duke Quantum Center, Duke University, Durham, NC, USA}
\affiliation{Department of Electrical and Computer Engineering, Duke University, Durham, NC, USA}
\affiliation{Dept. of Physics, Duke University, Durham, NC, USA}

\date{\today}

\begin{abstract}
Quantum networking can be realized by distributing pairs of entangled qubits between remote quantum processing nodes. Devoted communication qubits within each node can naturally interface with photons which bus quantum information between nodes. With the introduction of \ac{cqed} to enhance interactions between communication qubits and photons, advanced protocols capable of achieving high entanglement distribution rates with high fidelity become feasible. In this paper, we consider two such protocols based on trapped ion communication qubits strongly coupled to small optical cavities. We  study the rate and fidelity performance of these protocols as a function of critical device parameters and the photonic degree of freedom used to carry the quantum information. We compare the performance of these protocols with the traditional two-photon interference scheme, subjecting all protocols to the same experimentally relevant constraints. We find that adoption of the strong-coupling protocols could provide substantial distribution rate improvements of $30-75\%$ while maintaining the high-fidelities $\mathcal{F}\gtrsim99\%$ of the traditional scheme. 
\end{abstract}


\maketitle

\section{Introduction}
Remote, stationary qubits may be linked with ``flying" photonic qubits, generating \ac{re}~\cite{SimonPRL2003}. Optical transitions between the electronic states of many atoms or atom-like defects in solids make for a natural marriage between matter qubits and photons~\cite{moehring2007entanglement,RitterNature2012,BernienNature2013}, though a great bit of freedom remains in specific details of implementation. At the physical layer, we must determine which atomic and photonic \ac{dof} encode information as well as the mechanisms by which those \ac{dof} interact, paying close attention to error and loss channels inherent in that choice.  Abstracting to a link layer, we must also specify network topology and protocols to detect loss or error~\cite{beukers2024remote}. In this work, we explore this space of photonic qubits and entanglement protocols. Our analysis centers on qubits defined in the electronic levels of trapped atomic ions, though many of our considerations apply to other quantum platforms. 

Exploring trapped ion network designs makes for an immediately relevant case study. Trapped ions provide a promising platform for quantum memory, boasting excellent coherence times \cite{wang2017single}, vanishing state preparation and measurement errors \cite{an2022high, ransford2021weak}, and unrivaled high-fidelity one- and two-qubit gate operations \cite{harty2014high, loschnauer2024scalable, clark2021high}. The default local two-qubit interactions, mediated by the collective motion of co-trapped ions, enable information processing within the confines of an ion trap chip but \emph{do not} extend beyond the chip. Operations which transfer quantum information between chips are necessary for both modularized processor architectures \cite{monroe2013scaling, monroe2014large} and distributed quantum networks. 

There are two traditional approaches to \ac{re} generation, \textit{type-I}, which succeeds when exactly \textit{one} of two identical, weakly-excited atomic sources produces a photon, and \textit{type-II}, which succeeds upon simultaneous detection of \textit{two} photons from atomic sources excited with near-unit probability \cite{monroe2014large, luo2009protocols}. The type-I protocol requires an interferometrically stable optical path and is subject to fidelity limitation due to multi-photon excitation events. Therefore, trapped ion \ac{re} demonstrations have predominantly used type-II protocols. We treat type-II as the benchmark protocol in this work as well, omitting any further discussion of type-I.

Branching spontaneous decays in atoms naturally emit photons with \ac{dof} already entangled with the resulting atomic state, and the type-II protocol leverages this built-in entangled pair-production mechanism. This approach works across several photon \ac{dof} encodings; The earliest demonstration encoded qubits in frequency \cite{moehring2007entanglement}, the fastest demonstrations encoded in polarization \cite{stephenson2020high, jameson2024fast}, and the highest-fidelity demonstrations encoded in time-bins \cite{saha2024high}. Since the earliest experiments, \ac{re} success rates have improved by more than five orders of magnitude \cite{moehring2007entanglement, jameson2024fast}. Even so, the best average \ac{re} generation time is about 40 times slower than local entangling gates in trapped ion systems, limited by a squared dependence on historically poor photon collection efficiencies. This imposes a major architectural bottleneck to efficient distributed quantum computing \cite{hankin2024velociti}.

Optical cavities promise to enhance collection rates with the Purcell effect \cite{purcell1995spontaneous, kim2011efficient}, alleviating the \ac{re} bottleneck. Several groups have endeavored to integrate optical cavities with trapped ions \cite{sterk2012photon}, some nearly saturating upper bounds on quantum collection efficiency \cite{schupp2021Interface}, though an improvement over free-space collection with a high numerical aperture (NA) lens (e.g. \cite{carter2024ion}) remains to be seen. Purcell-enhanced photon collection is just one example from a family of behaviors described by \ac{cqed}. The amplified atom-light coupling underpinning Purcell enhancement also enables several alternative entanglement mechanisms which may be used to realize several other \ac{re} protocols \cite{duan2004scalable, cirac1997quantum, welte2017cavity, beukers2024remote}, the focus of this work.

For trapped ion networks, we often regard cavities merely as a solution to enhance photon collection. Here we ask whether access to similar cavities, refined to access the \ac{scr}~\cite{WaltherRPP2006}, may make alternative \ac{re} generation protocols viable in trapped ion systems. With practical constraints on cavity size and loss, we look to identify parameter regimes where these \ac{scr} protocols offer a competitive advantage over the conventional type-II procedure. We investigate optimal performance for two \ac{scr}-based protocols and compare their performance against the type-II approach. We consider three compatible photon modalities for each protocol and focus on limitations inherent to the protocols and cavity quality. 

The remaining sections are arranged as follows: \Cref{sec:background} reviews relevant results of \ac{cqed} and explores practical limitations of Purcell enhanced collection. \Cref{sec:qbs} outlines the photonic qubit encodings we consider. \Cref{sec:ExF} extends the \ac{cqed} analysis to accommodate external light fields and identifies two new mechanisms for ion-photon entanglement. \Cref{sec:protos} summarizes type-II entanglement and introduces two alternative single photon protocols, highlighting efficiency and fidelity limitations of each. \Cref{sec:results} compares achievable protocol fidelities and rates across encodings and protocols.

\section{Atomic Qubits in Cavities\label{sec:background}}

\subsection{Summary of Cavity QED\label{sec:cqed}}
CQED describes atom-light interactions when a particular privileged mode of the electromagnetic field in an optical cavity resonantly couples to two-level photon emitters \cite{harocheCavity1989}. We focus on the simplest example of a single atom spanned by a ground and excited state, $|\text{g}\rangle$ and $|\text{ex}\rangle$, interacting with a single bosonic cavity mode. We approximate unitary dynamics with a \ac{jch}, $\hat{\mathcal{H}}_{JC} = g(\hat{\sigma}_+\hat{a} + \hat{\sigma}_-\hat{a}^\dagger)$, where $\hat{\sigma}_-$ and $\hat{a}$ [$\hat{\sigma}_+$ and $\hat{a}^\dagger$] are the lowering [raising] operators for the atom and cavity field, respectively \cite{JaynesComparison1963}. The coupling rate $g$ represents coherent interaction strength.  

An excited atom placed in a dark cavity ($|\text{ex}, 0\rangle$) may emit a single photon into the cavity mode, resulting in the state $|\text{g}, 1\rangle$. In turn, the photon can be reabsorbed by the atom, leading to \textit{vacuum Rabi oscillations} at the rate $|2g|$ for resonant systems. In this ideal system, population never leaves this two-dimensional subspace because the total number of excitations $\hat{N} = \hat{a}^\dagger\hat{a} + \hat{\sigma}_+\hat{\sigma}_-$ is conserved  under the \ac{jch}: $[\hat{\mathcal{H}}_{JC},\hat{N}]=0$ \cite{scully1997quantum}. The maximum strength $g_o$ depends on the atomic transition dipole moment $\mu_{eg}$, photon frequency $\omega$, and the mode volume $V$ of the cavity, and is given by

\begin{equation}
    g_o = \frac{\mu_{eg}}{\hbar}\sqrt{\frac{\hbar\omega}{2\varepsilon_o V}},
\end{equation}

\noindent where $\varepsilon_o$ is the electrical permittivity of vacuum, achieved when the atom sits at the location of strongest electric field with its dipole moment aligned to the field polarization. In a more general case, an ion at a position $\vec{r}$ experiences coupling strength $g = g_o u(\vec{r})( \hat{\mu}_{eg}\cdot\hat{\epsilon})$, where $u$ is the position-dependent field amplitude normalized to a maximum of 1, $\hat{\mu}_{eg}$ is the unit vector of the transition dipole, and $\hat{\epsilon}$ is the field polarization unit vector. Mode volume is calculated as $V\equiv\int |u(\vec{r})|^2 d\vec{r}$. For a standing-wave Gaussian mode with waist $w_o$ between mirrors separated by length $\ell$, this evaluates to $V=\pi w_o^2 \ell / 4$. An ion along the cavity axis at a distance $z$ from the mode waist experiences a diverged beam with radius $w_z > w_o$ and field amplitude reduced by a factor $w_o/w_z$ from the maximum. We incorporate this reduction by considering an effective mode volume $\widetilde{V} = \pi w_z^2 \ell / 4$.

Realistically, system excitations leak from imperfect cavities or by spontaneous emission. The rate of electric field leakage from the cavity $\kappa = \pi\nu_F/\mathscr{F}$ depends on the free spectral range $\nu_F = c/2\ell$ (where $c$ is the speed of light) and cavity finesse $\mathscr{F}\approx -\pi/\ln\sqrt{1-\mathcal{L}}$, defined for round-trip intensity loss $\mathcal{L}$ \cite{saleh2019fundamentals}. For very low loss cavities ($\mathcal{L}\ll 1$), $\mathscr{F} \simeq 2 \pi / \mathcal{L}$ and $\kappa(\mathcal{L})\simeq \nu_F\cdot \mathcal{L}/2$. Excited atoms can also spontaneously emit into free-space instead of the privileged cavity mode, at a rate $\gamma$ which is proportional to the density of accessible free-space modes \cite{weisskopf1997berechnung}. We treat $\gamma$ as a constant intrinsic to the ion species, neglecting the solid angle of free space blocked by cavity mirrors which is typically small \cite{reiserer2015cavity}. In the Linblad formalism \cite{gorini1976completely}, we model damped \ac{jch} evolution by including two collapse operators $\hat{L}_{C} = \sqrt{2\kappa}\hat{a}$ and $\hat{L}_{S} = \sqrt{2\gamma}\hat{\sigma}_-$, representing cavity loss and spontaneous emission, respectively. We only observe vacuum Rabi oscillations when $g$ outpaces leakage  $g>\kappa, \gamma$, the parameter range defining the \ac{scr}. Without cavity coupling, excited atom population decays exponentially with time constant $\Gamma^{-1}$, where $\Gamma \equiv 2\gamma$ is the \ac{fwhm} of the atomic absorption line.

The transition dipole moment $\mu_{eg}$ is related to $\Gamma$ by

\begin{equation}
    \mu_{eg}^2 = 3\pi\varepsilon_o\hbar\Gamma c^3/\omega^3 \times R_{br} \times W^2_{eg},
\end{equation}

\noindent where $R_{br}$ is the branching ratio from the excited state to the desired ground-state manifold, and $W_{eg}$ captures the dipole overlap between excited and ground states, determined by the Wigner-Eckart theorem \cite{James1997QuantumDO, king2008angular}.

\subsection{Collection from Leaky Cavities}
Transmitted cavity photons emerge in a single mode, more readily coupled into a single mode fiber than free-space emissions. Transmissions comprise only a fraction of cavity decays, so we subdivide $\kappa = \kappa_L + \kappa_R + \kappa_B$, denoting leakage due to left- and right-side transmission (assuming a two-port cavity) and  \textit{bad losses} $\mathcal{L}_B$. Bad-loss is a catch-all term for uncollectable scattering $\mathcal{L}_s$ and absorption losses $\mathcal{L}_a$. There is no physical significance behind terming a side left or right, but we assume we collect transmissions from the left port. If an excitation pulse initializes \footnote{Realistically, the pulse will not be instantaneous, however it must be much faster than $\gamma$ to avoid sick decays \cite{wang2020direct,gorshkov2013dissipative}} the atom-cavity system $|\psi(t)\rangle$ to $|\text{ex},0\rangle$ at $t=0$, the probability of photon output through time $s$ accumulates as $P(s)=2\kappa_L\int_0^s |\langle \psi(t)| \text{g},1\rangle|^2dt$. 

$|\psi(t)\rangle$ evolves according to the damped \ac{jch}, leading to an open-systems generalization of vacuum Rabi oscillations. Ground and excited state amplitudes exhibit exponentially decaying oscillations with modified frequency $g' = \sqrt{g^2 - (\kappa-\gamma)^2/4}$ and averaged-linewidth decay constant $K \equiv (\kappa + \gamma)/2$ \cite{cui2005quantum}. For $s\gg K^{-1}$, we may approximate $P(s)$ with its long-time limit $P_1\equiv\lim_{s\rightarrow\infty}P(s)$ as

\begin{equation}\label{eqn:p1}
    P_1 = \underbrace{\frac{g^2}{g^2 + \kappa\gamma}}_{\eta_c}\cdot \underbrace{\frac{\kappa_L}{\kappa + \gamma}}_{\eta_{ex}}.
\end{equation}

We factorize $P_1$ as a product of \textit{coupling} and \textit{extraction} efficiencies, $\eta_{c}$ and $\eta_{ex}$, respectively. The coupling efficiency is often reported as $C/(C+1)$ in terms of the cooperativity $C\equiv g^2/\kappa\gamma$. $C$ is also known as the Purcell factor, the multiplicative increase in the density of states in a cavity over free space which leads to enhanced emission rates \cite{purcell1995spontaneous, barnes2020classical}. Note, it's common to see total collection and coupling efficiencies conflated $P_1\simeq \eta_{c}$ \cite{law1997deterministic}, which is an accurate simplification in the regime where $\kappa_L\approx\kappa\gg\gamma$ when transmission is the dominant loss mechanism. Commonly studied trapped ion transitions feature short excited state lifetimes and deep-blue transitions which are more readily scattered and absorbed at mirrors so, for our purposes, $\eta_{ex}$ should not be ignored. 

High cooperativity alone is insufficient to guarantee efficient $P_1$. Rather, we require transmission simultaneously fast enough to dominate unusable losses and slow enough to maintain high-cooperativity. This describes the leaky cavity regime, where

\begin{equation}\label{eqn:leaky}
    \kappa_L , \frac{g^2}{\kappa} \gg \gamma.
\end{equation}

In some cases, one excited state might decay to several ground states. If multiple decay pathways couple to different modes in the same cavity, we must substitute $g\leftarrow\widetilde{g} \equiv \sqrt{\sum_i g_i^2}$, where the $g_i$ are calculated for the individual transitions, to predict total collection with $P_1$. 

\subsection{Sub-millimeter Cavities for Collection\label{sec:purcell}}
\begin{figure}
    \centering
    \includegraphics[width = 0.7\linewidth]{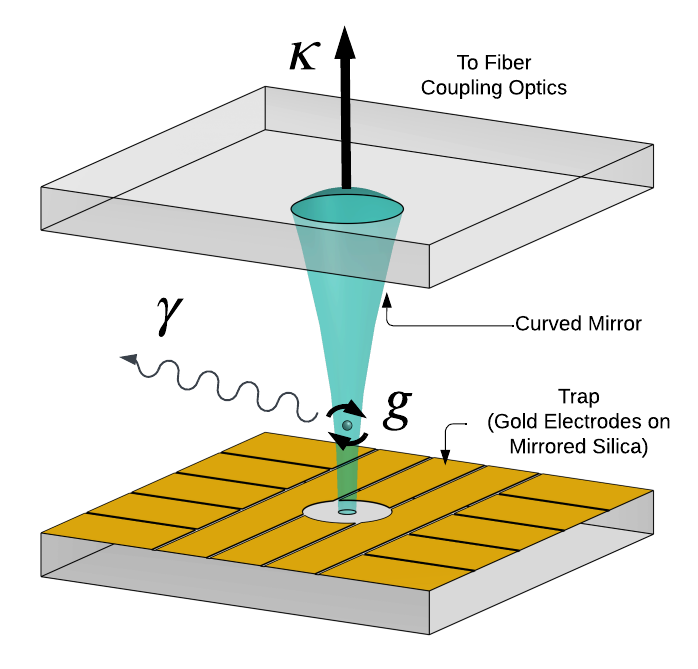}
    
    \caption{Integrated optical cavity design composed of a curved mirror etched into a glass substrate suspended above a micro-fabricated surface-trap where small gaps in the electrodes reveal flat mirrored surfaces.}
    \label{fig:micro_cav}
\end{figure}

Consider a surface-trap integrated with a small optical cavity as in \cite{kim2011efficient, van2016integrated}, composed of a flat mirror embedded in the trap surface with a spherical mirror with radius of curvature $R$ centered above the trap (\Cref{fig:micro_cav}). 

When $\ell<R$, this configuration isolates a Gaussian mode with waist located at the flat mirror and a Rayleigh range $z_R = \sqrt{R\ell - \ell^2}$. Planar traps confine ions some finite height $h_{ion}$ above the surface, typically $\sim70\unit{\micro\meter}$ \cite{revelle2020phoenix}, preventing atoms from residing at the cavity mode waist. The relationships between $R,\ell$ and $h_{ion}$ provide useful freedom in determining cavity performance.

The tightest achievable focus at the ion obeys $\min(r_{ion}) = \sqrt{2\lambda h_{ion}/\pi}$, independent of $R$, attained for cavity length $\ell_o$ satisfying $z_R = h_{ion}$.  If $R$ is fixed, $C$ is maximized at $\ell_o$. This $\ell_o$ is not exactly the same as the $\widetilde{V}$-minimizing length (where $g$ is maximized), though it serves as a good approximation due to a rapid de-focusing of $r_{ion}$ away from $\ell_o$. An order-of-magnitude estimate of this minimum volume is $\min(\widetilde{V})\sim\lambda h_{ion}R/2$, most accurate when $R\gg h_{ion}$, implying that tightly curved mirrors are a crucial prerequisite for efficient photon collection.

In principle, different approaches to cavity integration can confine the ion at the mode waist \cite{schupp2021Interface, david2023progress} allowing for $\widetilde{V}$ arbitrarily close to zero. Even so, vanishing waists increase sensitivity to ion micromotion and imply rapid mode divergence, requiring high-numerical aperture (NA) curved mirrors and optomechanical stability. Under the constraint of a minimum $w_{ion}$ or maximum $\text{NA}$, small-$R$ mirrors are still favorable when targeting strong coupling. 

\begin{figure}
    \centering
    \includegraphics[width = 0.9\linewidth]{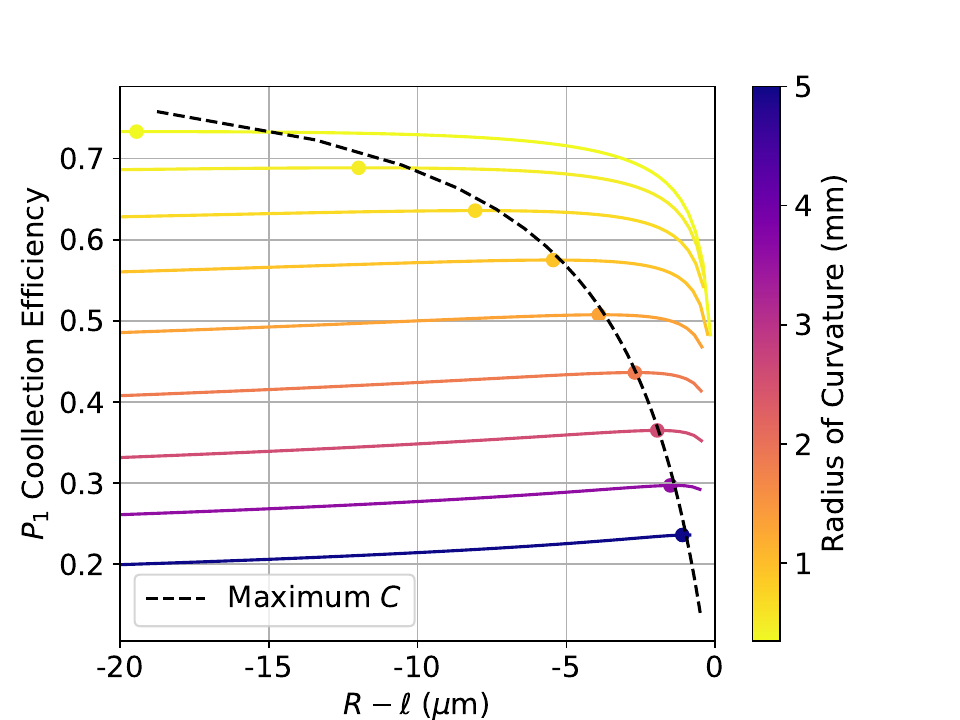}
    
    \caption{Simulated collection efficiencies from a  Ba-like transition with $\lambda=493\unit{\nano\meter}$ and $\mu = 2.34e \text{a}_o$ for varied mirror $R$ and fixed $\mathscr{F}=4000$ and $h_{ion}=70\unit{\micro\meter}$. Smaller $R$ correspond to greater maximized $P_1$ (circles) and decreased sensitivity to cavity length. The dashed line shows the coopertivity maximizing length where $z_R=h_{ion}$, which predicts the max-$P_1$ lengths to reasonable accuracy.}
    \label{fig:perf_range}
\end{figure}

To illustrate the impact of reducing mirror radius, we model collection from a bad-loss-free emission cavity built with mirror radii ranging from $R=5\unit{\milli\meter}$ down to $300\unit{\micro\meter}$ in \cref{fig:perf_range}. Maximum collection efficiency improves as $R$ decreases. Efficiency is also less sensitive to deviations from the optimal length (marked) for the smaller cavities. Finally, we note that the the optimal lengths for large-$R$ cavities approach the hemispherical-limit $\ell\approx R$, where rapidly diverging modes require high-NA mirrors.

Improved efficiency for small-$R$ assumes mirror sizes can be decreased without compromising surface quality. Scattering losses from mirror surfaces grow approximately as $\mathcal{L}_s \approx 1 - \exp\big[-(\frac{4\pi\sigma}{\lambda})^2\big]$ \cite{bennett1992recent}, where $\sigma$ is the RMS surface roughness integrated over spatial frequencies lower than the inverse wavelength $\lambda^{-1}$\cite{harvey2012total}. The super-polishing techniques traditionally used to fabricate mirrors with surface roughness $\sigma<1\text{\AA}$ do not extend well to $R\lesssim 5\unit{\milli\meter}$ \cite{jin2022micro}. To meet the challenge of producing smooth mirrors with $R<1\unit{\milli\meter}$, several groups have turned to laser ablation of glass surfaces. A spatially filtered high-power laser pulse at $10.6\unit{\micro\meter}$ can ablate a crater with near-spherical geometry near the nadir \cite{takahashi2014novel}, and subsequent lower-power melting pulses allow surface tension to smooth imperfections \cite{nowak2006efficient}. These mirrors may be fabricated onto a bulk fused-silica substrate \cite{van2016integrated} or directly onto the tip of an optical fiber \cite{takahashi2014novel, hunger2010fiber, brandstatter2013integrated}. With these methods, mirrors with $R\lesssim400\unit{\micro\meter}$ and $\sigma_{RMS}\lesssim 2\text{\AA}$ may readily be fabricated. 

\subsection{Reaching the Strong Coupling Regime}
For most atomic dipole transitions, spontaneous emission cannot be changed appreciably from $\gamma\sim 2\pi \times 10\unit{\mega\hertz}$, setting a minimum requirement for reaching the \ac{scr}. In particular, with the minimum sufficient coherent coupling rate $g\simeq\gamma$, we need $\kappa\leq\gamma$. With an $\ell=500\unit{\micro\meter}$ cavity, this requires $\mathscr{F}\geq 15,000$. This describes the threshold to the \ac{scr} where $C=1$, but higher cooperativity is possible.  In principle, that small cavity may reach $g\simeq 2\pi\times 65\unit{\mega\hertz}$ in the optical frequencies (e.g. $\lambda=493\unit{\nano\meter}$ for $\text{Ba}$) with a mode waist of $w_o=3\unit{\micro\meter}$, in which case $C\gtrsim40$. $C$ can further be improved by increasing $\mathscr{F}$ or moving to cavities with smaller $w_o$.

\section{Ion-Photon Entanglement}\label{sec:qbs}
Single photons can store quantum information in their internal \ac{dof} such as polarization, frequency, and timing. These three modalities may be generated from Purcell-enhanced emitters, entangling photonic \ac{dof} with electronic state of the source atom \textit{ab initio}. Quantum information can also be stored in photon intensity and spatial modes, respectively referred to as as \textit{Fock state} and \textit{dual-rail} encoding. Photon losses inherent in any transmission channel severely limit the practicality of Fock state qubits in lossy systems, so we omit their consideration from further discussion. Atoms do not directly emit into dual-rail photons, but one can convert polarization, frequency, or time-bin qubits to dual-rail qubits which proves to be useful for state detection. Here, we outline mechanisms which generate ion-photon Bell states. We also summarize potential error sources and implementation challenges specific to each modality. Finally, we give a brief overview of methods to convert these photonic qubit variants to dual-rail encodings. 

\subsection{Polarization Qubits} If a single excited state decays to two ground states with different total-electronic magnetic quantum numbers $m_j$, the emitted photon polarization will be entangled with $m_j$, see \cref{fig:simple_exc}a.  Ideal optical cavities support two orthogonal polarization modes, so both decay pathways may couple to the same cavity. In free-space, the splitting ratio between these two paths follows from the relative strength of $\mu_{i}$, where $i\in\{1,2\}$ indexes the transitions. In a cavity, however, the alignment $\hat{\mu}\cdot\hat{\epsilon}$ modulates coupling strength. For an even superposition of cavity-generated photon states, $g_o^{(i)}(\hat{\mu}_i\cdot\hat{\epsilon_i})$ should be made constant between the two transitions. Control of the angle between the external magnetic field (quantization axis) and the cavity axis provides some freedom to compensate unequal $g_o^{(i)}$. Resolvable frequency differences between the two transitions $|\omega_1-\omega_2|>\Gamma$ produce hyper-entangled photons which will decohere if the environment inadvertently measures photon frequency. To protect polarization qubits from these channels, decays which branch to different $L_j$-manifolds or hyperfine levels with large frequency differences should be avoided. As a consequence of the encoding, atomic qubit states are magnetic-field sensitive Zeman states, so a weak quantization field is critical to keeping the $\omega_i$ unresolved. Even so, the atomic qubit should be transferred to a more stable splitting (e.g. hyperfine) after \ac{re} to prevent decoherence by magnetic field fluctuations during storage.

For a well defined direction of propagation, the 2-dimensional polarization state-space accommodates exactly one qubit, permitting no buffer for rotation errors. Any misalignment of polarization axes creates a channel for undetectable bit-flip errors (as opposed to a detectable erasure). This applies to interfaces between fibers as well as the wave-plates and polarizing beam-splitters for qubit manipulation and detection. Misalignment between the cavity and quantization axes also introduces crosstalk during generation. Furthermore, most transmissive media exhibit some degree of birefringance, imposing asymmetry on the polarization modes. \Ac{pdl} distorts and mixes qubit amplitudes. Anisotropy in refractive index will temporally separate modes during transmission, an effect known as \ac{pmd}, entangling the qubit state with the time \ac{dof} as well as elongating photons and limiting the maximum bit-rate. The classical telecom sector avoids polarization as a viable channel for multiplexing long-haul signals because of these birefringance-induced effects \cite{ramaswami2009optical, agrawal2012fiber}. While polarization qubits offer the opportunity for the fastest RE distribution, retaining high fidelity is a challenge due to the stringent requirements on polarization maintenance.


\subsection{Frequency Qubits}
An atomic decay branching to two ground states with resolved energies emits a frequency-binned qubit entangled with the atomic state. Here, we focus on decays to ground states with identical $m_j$ to avoid the challenges of asymmetric cavity coupling and birefringance.  With identical transition dipole alignment $\hat{\mu}_i$, balanced splitting requires branches where $|\mu_{1}|=|\mu_{2}|$. 

\begin{figure}
    \centering
    \includegraphics[width = 0.9\linewidth, trim={0 0cm 0 0cm}, clip]{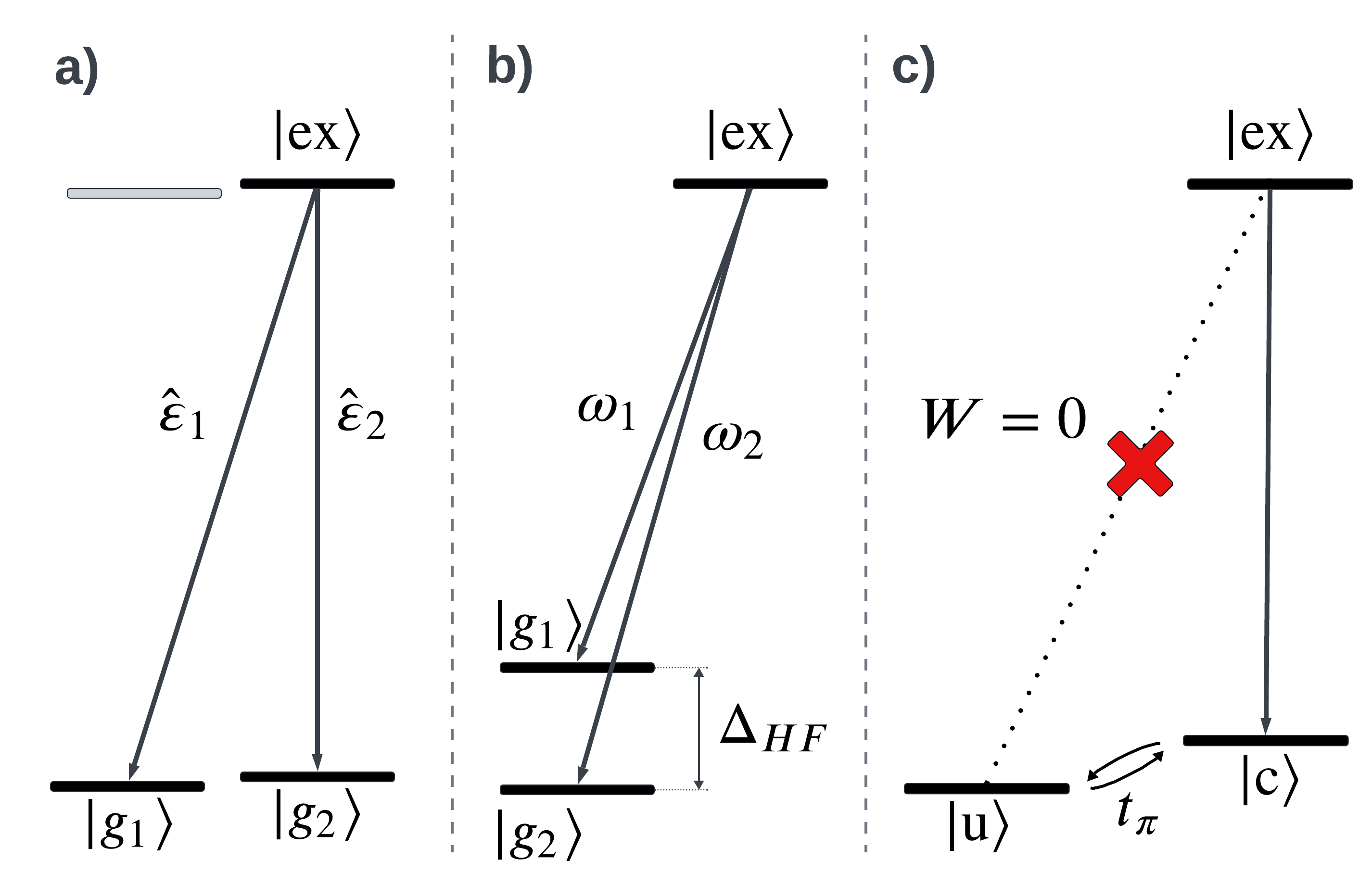}
    
    \caption{Simplified level diagrams of ion-photon entangled pair production. (a) Production of a polarization qubit where decay to states with different $m_j$ produces photon states with different $\hat{\varepsilon}_i$. (b) Frequency qubit production where decays to levels with hyperfine splitting $\Delta_{HF}$ produces photon states with  with different $\omega_i$. (c) Time-bin qubit production where decays to $|\text{c}\rangle$ are allowed and decays to $|\text{u}\rangle$ are not. After excitation of $|\text{c}\rangle$ and emission populating $|\text{e}\rangle$ time-bin, a resonant $\pi$-pulse inverts population between $|\text{u}\rangle\leftrightarrow|\text{c}\rangle$ so that a second excitation can populate $|\text{l}\rangle$ time-bin.}
    \label{fig:simple_exc}
\end{figure}

Decays to magnetic-field insensitive ``clock" transitions in the hyperfine ground-state manifold provide a compelling option \cite{moehring2007entanglement}. Beyond outstanding frequency stability, these $\Delta_{HF}\sim10\unit{\giga\hertz}$ splittings are easily resolved by atoms with $\Gamma\sim20\unit{MHz}$ but also co-propagate through most passive optics without discrimination.  Non-lossy qubit rotations may, in principle, be implemented with pulse-shaped electro-optic modulation techniques \cite{lukens2016frequency, lu2018electro}. While not imperative, we will assume hyperfine-split frequency qubits in the remaining discussion, see \ref{fig:simple_exc}b.


To enhance both decay channels, optical cavities must simultaneously resonate with both frequencies, constraining cavity design. $\Delta_{HF}$ must either fit within a cavity linewidth or match the spacing between adjacent cavity resonances. The former requires $2\kappa\gg\Delta_{HF}$, leading to low $C$ for most ion-based systems. The latter requires $\Delta_{HF}=\nu_F$, implying a cavity length of $\ell\approx1.5\unit{\centi\meter}$ for a $10\unit{\giga\hertz}$ splitting. As explored in \cref{fig:perf_range}, the advantages of cavity-enhanced collection are diminished at large scales,
and the resulting collection efficiencies would not be competitive with lens-based approaches \cite{carter2024ion}.

Also, due to the differing $\lambda_i$, the standing wave patterns $u_i(\vec{r})$ will not match. Care must be taken to situate the ion at an $\vec{r}$ where $u_1(\vec{r})\approx u_2(\vec{r})$ and $u_{1,2}(\vec{r})\approx 1$ to maintain strong, balanced emission. Finally, large splittings imply rapid accumulation of qubit phase during propagation, meaning that small drifts in path length can dephase the qubit. Tolerance varies with fidelity requirements, but for high-fidelity operation with $1-\mathcal{F} < 10^{-3}$ we estimate a maximum acceptable path-length uncertainty of $\sigma[\Delta Z]\lesssim 100\unit{\micro\meter}$. 

\subsubsection*{Optical Qubits}
Decays to states with even larger frequency splittings of $\Delta_O\gtrsim 100\unit{\tera\hertz}$, often seen between different $L_j$-manifolds, are called \textit{optical} qubits and could be considered an alternative type of frequency qubit. The large frequency difference between qubit states might impose optical incompatibilities such as dispersion or non-single-mode behavior in fiber or unequal loss in coatings. Furthermore, photonic qubit rotations for basis states with such dissimilar energies requires quantum frequency conversion with power-intensive nonlinear techniques \cite{lu2023frequency}. Tracking the relative phase evolution requires optical clocks which are neither as accurate nor as robust as the microwave clocks required for tracking hyperfine splittings. Finally, distributing these $\Delta_O$-scale control signals requires interferometrically-stability with $\sigma[\Delta Z]\ll c/\Delta_O \sim 1 \unit{\micro\meter}$ path length tolerances at distance. Due to these technical challenges, we will not consider optical qubits in this work. 

\subsection{Time-Bin Qubits}
A time-bin qubit is composed of a pair of temporally-resolved identical wavepackets called the early $|\text{e}\rangle$ and late $|\text{l}\rangle$ time bins, separated by a time interval $\Delta t$. The source exclusively populates one bin or the other, allowing both possibilities in superposition if the source states are initialized in superposition. The same atomic decay pathway populates both bins, so a time-bin emitter cavity only needs to couple to one transition. This allows us to select un-split transitions ($W_{\text{ex,c}} = 1$) often implying stronger interactions.

Our ideal atomic source consists of two states $|\text{u}\rangle$ (uncoupled) and $|\text{c}\rangle$ (coupled) in the ground state manifold and one excited state $|\text{ex}\rangle$, where $|\text{ex}\rangle\leftrightarrow|\text{c}\rangle$ is a strong cycling transition enhanced by the cavity. After excitation of $|\text{c}\rangle\rightarrow|\text{ex}\rangle$, a photon emission reliably restores the atom state to $|\text{c}\rangle$ within some decay window longer than the cavity-enhanced decay time, $s\gg K^{-1}$. Importantly, $|\text{u}\rangle$ remains untouched in this excitation loop.  To generate a time-binned photon, we initialize the source atom in $\alpha|\text{c}\rangle + \beta|\text{u}\rangle$. These $\alpha,\beta$ are arbitrary state amplitudes which should be identically $1/\sqrt{2}$ for Bell-state production. One excitation loop generates the early time-bin with amplitude $\alpha$, a $\pi$-pulse inverts atomic qubit populations, and a subsequent excitation loop generates the late time-bin with amplitude $\beta$. In total, this excite-flip-excite sequence imprints the the initial atomic qubit superposition onto the joint ion-photon state $\alpha|\text{u}, \text{e}\rangle + \beta|\text{c},\text{l}\rangle$. The separation between time-bins is then $\Delta t = s + t_\pi$, where $t_\pi$ is the $\pi$-time of the interceding rotation.


The three-part excite-flip-excite sequence generates longer qubits, spanning $2s + t_\pi$ in time, as opposed to the one-time excitation procedures which span only a single collection window $s$. Long photons cannot pack as densely onto optical fiber, lowering maximal bitrate for time-binned qubits. Ultra-fast pulse techniques could reduce $t_\pi$ \cite{mizrahi2014quantum, campbell2010ultrafast}, but the two emission windows are unavoidable. On the other hand, by avoiding the storage of information in polarization or frequency dimensions, environmental interaction with those \ac{dof} cannot decohere qubits. For example, bin frequencies are identical and the phase accumulated during propagation is global, so slow path length fluctuations won't dephase time-binned qubits. Only environmental conditions which can meaningfully drift between the two time bins can impose asymmetries on the bins. For ions, we expect $\Delta t\lesssim 1\unit{\micro\second}$, protecting qubits from most thermal and mechanical fluctuations with much longer timescales \cite{bersin2024development}. 

\subsection{\label{sec:Dual_Rail}Dual-Rail Qubits and X-Basis Measurements}
\begin{figure*}
\includegraphics[width=0.75\linewidth]{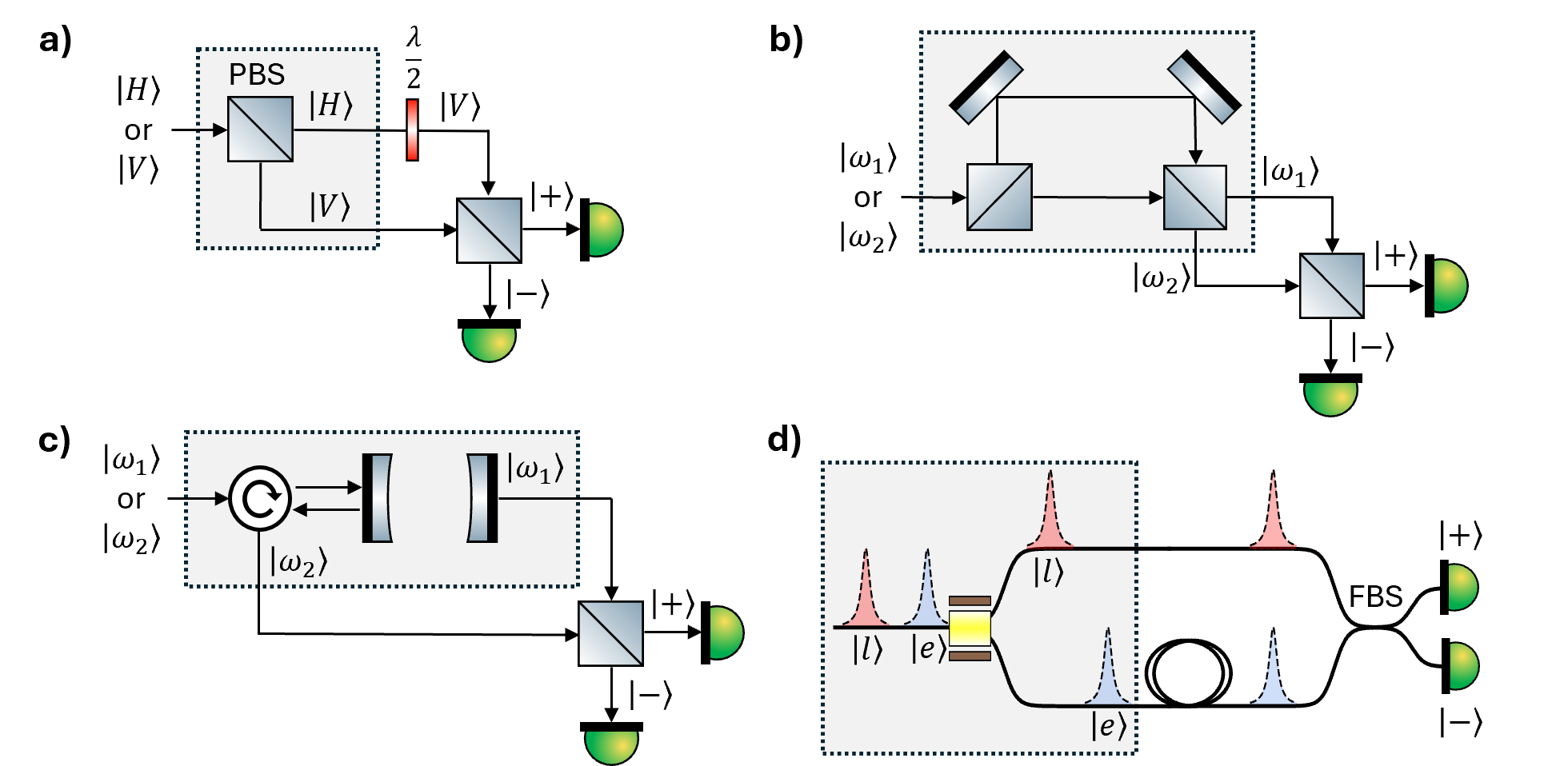}
\caption{\label{fig:dummy} Configurations for $X$-basis measurements on photonic qubits encoded in polarization (a) , frequency (b and c), and time-bin (d) qubits. All measurements begin with a conversion to dual-rail encoding (light grey box), potentially followed polarization or time-bin concealment, and finally interference at a \ac{npbs} to complete the $X$-basis measurement. The single photon detectors could be placed directly at the outputs of the dual-rail encoding section, bypassing the interference,  to complete a measurement in the computational basis as well.  }
\end{figure*}

The ability to convert polarization, frequency, and time-bin qubits into dual-rail encoding is used for measurement or filtering photons by state. Dual-rail qubits, where the photon can take one of two spatial modes \cite{RevModPhys.79.135}, provide a convenient format for photonic qubit detection given the availability of high performance single photon detectors today \cite{semenov2001quantum, marsili2013detecting}, some with photon-number resolving capabilities \cite{cahall2017multi, cheng2023100}. Polarization qubits can readily be converted to dual-rail by using a polarizing beam-splitter (\cref{fig:dummy}a). Frequency qubits can be converted to dual-rail using an Mach-Zehnder interferometer featuring a delay line in one branch to cause each qubit frequency to constructively interfere in a distinct output port (\cref{fig:dummy}b). Alternatively, a cavity resonant with only one of the frequencies will couple that resonant light into the transmitted mode while the other frequency state reflects (\cref{fig:dummy}c). Time-bin qubits can be converted to dual-rail with near-unit efficiency using a fast switch synchronized with the time-bins (\cref{fig:dummy}d). 


Dual rail qubits with computational ($Z$) basis defined by spatial modes $|\text{a}\rangle$ and $|\text{b}\rangle$ are also readily measured in the $X$-basis $|\pm\rangle_{(DR)}\equiv (|\text{a}\rangle\pm|\text{b}\rangle)/\sqrt{2}$. A 50:50 \ac{npbs} maps photons input modes $|\text{a}\rangle$ and $|\text{b}\rangle$ to the output modes $|\text{c}\rangle$ and $|\text{d}\rangle$ via the beam-splitter transform $|\text{a}\rangle\rightarrow (|\text{c}\rangle + |\text{d}\rangle)/\sqrt{2}$ , $|\text{a}\rangle\rightarrow (|\text{c}\rangle -|\text{d}\rangle)/\sqrt{2}$. This means that a photon detection at output mode $|\text{c}\rangle$ ($|\text{d}\rangle$) implies a pre-\ac{npbs} state of $|+\rangle_{(DR)}$ ($|-\rangle_{(DR)}$). 

This simple measurement only works if the wavepackets in either rail are identical in all \ac{dof} besides the spatial mode. Since the conversion techniques above segregate photons using differences in their polarization, frequency or time-bin \ac{dof}, inadvertent measurement of that original encoding collapses dual-rail superpositions, spoiling the $X$-basis measurement. To perform uncorrupted $X$ measurements, the original encoding must be erased from the photon state. For polarization qubits, a half-waveplate can rotate a $|\text{H}\rangle$ polarized wavepacket in one rail to match the $|\text{V}\rangle$ wavepacket in the other rail. Likewise, a delay line in the $|\text{e}\rangle$-bin rail will allow $|\text{e}\rangle$ and $|\text{l}\rangle$ wavepackets to temporally overlap at the beamsplitter. For frequency qubits, fast single-photon detectors with timing jitter much shorter than the inverse of the frequency difference obscure frequency information by virtue of the Fourier transform. If sufficiently fast detectors aren't available, frequency conversion using a short pump pulse has also been employed to broaden photon bandwidths and erase frequency information \cite{de2012quantum}. 

\section{Interacting with External Fields}\label{sec:ExF}
So far, we have focused on how emitter cavities improve out-coupling efficiency. $\hat{\mathcal{H}}_{JC}$ is Hermitian and governs unitary, and hence reversible, exchange. The concept of reversing the emission process to facilitate a near-deterministic photon absorption is well-established as a potential approach to quantum-state transfer \cite{cirac1997quantum}. Note, however, that the approach in \cite{cirac1997quantum} suffers from unfaithful transmissions when $\kappa_B$ or $\gamma$ are non-negligible and provides no simple means for post-selection to trade-off rate for improved fidelity. Here, we consider two alternative mechanisms where an external photon interacts with the receiver system as a scattering center. The scattered photon allows for some post-selection. 

Unlike cavity modes, external light sources do not necessarily have discretized spectra nor do they organize as stationary standing waves. We must treat external fields as collections of modes with continuous frequency spectra. Continuing with the example of a Fabry-Perot cavity, the left and right ports each couple to different continua. If we assume the leakage rates $\kappa_L$ and $\kappa_R$ remain constant over the frequency range of incident fields, input-output formalism of \cite{gardiner1985input, carmichael1993quantum} apply. The spectral amplitudes of the external field long-before ($\widetilde{a}_{\text{in},j}(\omega)$) and long-after ($\widetilde{a}_{\text{out},j}(\omega)$) interaction with the scatterer obey the input-output relationship

\begin{equation}
\widetilde{a}_{\text{out},j}(\omega) + \widetilde{a}_{\text{in},j}(\omega) = \sqrt{2\kappa_j}\widetilde{a}(\omega), j\in\{\text{L,R}\},
\end{equation}

\noindent where $\widetilde{a}(\omega)=\int_{-\infty}^{\infty}\hat{a}(t)e^{i\omega t\, dt}$ is the Fourier transform of the annihilation operator $\hat{a}(t)$ for the intra-cavity mode in the Heisenberg picture. Note, the $\widetilde{a}_{\text{out},j}(\omega)$ and $\widetilde{a}_{\text{in},j}(\omega)$ are \emph{not} standard annihilation operators, but rather flux operators with units of $\sqrt{\unit{\hertz}^{-1}}$. The result of scattering off of the ion-cavity system is expressed as a ratio of these flux operators. Assuming light impinges only from the left port, we define $r(\omega)\equiv \hat{a}(\omega)_{\text{out,L}}/\hat{a}(\omega)_{\text{in,L}}$ and $t(\omega)\equiv \hat{a}(\omega)_{\text{out,R}}/\hat{a}(\omega)_{\text{in,L}}$, the frequency-dependent reflection and transmission coefficients of the optical field incident on the cavity. The specific frequency dependence follows from the time dynamics of $\hat{a}(t)$ which may be determined by solving the Schrodinger equation for the damped-cavity system with a driving term $\hat{a}_{\text{in}, L}(t) = \frac{1}{2\pi}\int_{-\omega_o}^\infty \widetilde{a}_{in}(\omega)e^{-i\omega t}d\omega$ \cite{raymer2024duan}, resulting in

\begin{equation}\label{eqn:ref}
    r(\omega) = 1 - \frac{2\kappa_L(i\Delta_a + \gamma)}{(i\Delta_c + \kappa)(i\Delta_a + \gamma) + g^2},
\end{equation}

\begin{equation}\label{eqn:trans}
    t(\omega) = \frac{2\sqrt{\kappa_L\kappa_R}(i\Delta_a + \gamma)}{(i\Delta_c + \kappa)(i\Delta_a + \gamma) + g^2},
\end{equation}

\noindent where $\Delta_c = \omega-\omega_c$ and $\Delta_a = \omega-\omega_a$ are the detunings of the cavity and atom from the incident light at $\omega$. This is a standard result, often derived in the limit of weak, slowly-varying drives \cite{waks2006dipole, hu2008deterministic}, but the result actually holds more generally \cite{raymer2024duan}, including the faster-drives we consider here. 


For the \ac{scr} schemes in this work, emitted photonic qubits scatter from a second receiver cavity containing an atom of the same species as the emitter atom. The outcome of the scattering process depends on both the state of the incoming photon and the state of the atom in the receiver cavity. The transition $|\text{c}\rangle\leftrightarrow |\text{ex}\rangle$, where $|\text{c}\rangle$ is one of the receiver qubit states, couples strongly to the cavity. Importantly, the transition from the other qubit state $|\text{u}\rangle$ does not couple. This is naturally the case for the time-bin qubit. For polarization and frequency qubits, we relabel one of the ground states as $|\text{c}\rangle$ in accordance with the state of the incoming photon and calculate the coupling strength $g_c$ as in \cref{sec:background}. By design, the other ground state, now called $|\text{u}\rangle$, should not couple $g_u\approx 0$ due to selection rules or large detunings from the incident photon. By toggling the state of the atom, we can switch between the coupled and uncoupled transmission (or reflection) spectra. We narrow our attention to the behavior for resonant light $\Delta_a=\Delta_c=0$ in two specific cavity configurations.

\subsection{Dipole Induced Transparency}
Free from the influence of an atom, reflections from a cavity may be entirely extinguished when that cavity is critically-coupled (\emph{critical-coupling} is a term from classical resonator optics and not related to the coherent coupling rate $g$). Physically, this requires destructive interference between two types of reflections: \emph{prompt reflections}, where the photon reflects from the input mirror without ever entering the cavity, and \emph{leakage} photons which circulate in the cavity but re-emerge back out the input-side mirror \cite{black2001introduction}. We consider cavities with low bad losses $\mathcal{L}_B\ll1$ where the critical coupling condition reduces to the description of a \emph{balanced}-cavity with $\kappa_L=\kappa_R$. In such a balanced configuration, \cref{eqn:trans} implies

\begin{equation}\label{eqn:dit}
    t^\circ_i = \frac{2\kappa_L}{\kappa}\frac{1}{1+C_i}, \quad i\in\{u,c\},
\end{equation}

\noindent where we use the $\circ$ superscript to indicate validity for resonant light and the cooperativity $C_i=g_i^2/\kappa\gamma$ has been defined identically to  \cref{sec:background}, evaluating to $C_{u} = 0$ for the uncoupled-state atom. The atomic state modulates transmission, an effect known as \ac{dit} \cite{waks2006dipole}. $|\text{u}\rangle$  atoms preserve the strong transmission of the bare balanced cavity, but $|\text{c}\rangle$ atoms suppress transmission by a factor $(1+C_c)^{-1}$. For large $g_c$, we might interpret the extinction as the resonant dip between normal modes of width $\kappa + \gamma$ split by $2g_c$ when the modes resolve $2g_c > \kappa+\gamma$, but the effect is actually more general. Extinction ratio only depends on $C_c$, so the effect persists when $\kappa \ll \frac{g^2_c}{\gamma}$ which does not necessarily require resolved splitting. Due to \ac{dit}, the photon is transmitted with high probability when the receiver atom is in $|\text{u}\rangle$  and reflected with high probability when the atom is in $|\text{c}\rangle$.

\subsection{Controlled Phase-Flips}\label{cpf}
Consider instead a cavity biased to leak exclusively from the input port, $\kappa_R\approx 0$, what we will call an \emph{imbalanced} cavity \footnote{Such cavities may also be called \textit{over-coupled}, but we don't use this terminology again to avoid confusion with coherent coupling $g$}. Given this imbalance, prompt reflections and leakage do not completely interfere in the left-side output mode. From equation \cref{eqn:ref}, we get

\begin{equation}\label{eqn:cpf}
    r_i^\circ = 1 - \frac{2\kappa_L}{\kappa}\frac{1}{1+C_i},\quad i\in\{u,c\}.
\end{equation}

\noindent We find that reflection from uncoupled cavities induces a $\pi$-phase shift $r_u^\circ\simeq-1$ if transmission dominates bad losses $\kappa_L\simeq\kappa\gg\kappa_B,\kappa_R$. With the atom in $|\text{c}\rangle$ , however, that phase-shift disappears $r_c^\circ\approx 1$, provided that $C_c\gg 1$. For less-cooperative cavities or those with non-negligible bad-losses, reflection intensity may reduce $|r_i|<1$, but the phase-flip of $\pi$ will remain exact for resonant photons (barring extremely poor cavity performance with $2\kappa_L<\kappa$). We consider this a \ac{cpf} where the outgoing photon acquires a phase-shift corresponding to the receiver atom state. 
\section{Entanglement Protocols}\label{sec:protos}
We consider three RE generation protocols in which photonic qubits shuttle quantum information between two remote atoms each housed in a stationary optical cavity. Cavities are either emitters, optimized for efficient photon collection, or receivers, configured to exhibit either the \ac{dit} or CPF effects outlined in \cref{sec:ExF}. Each protocol can be realized using one of the three photonic qubit modalities in \cref{sec:qbs}; we describe protocols using time-bin qubits for simplicity, and the analysis for the other qubit types follows by analogy.

We operate emitter cavities to probabilistically produce a maximally-entangled ion-photon state $\frac{1}{\sqrt{2}}\big(|\text{u,e}\rangle + e^{i\phi}|\text{c,l}\rangle\big)$. The relative phase $\phi$ is determined by Clebsh-Gordon coefficients, qubit splittings, and path lengths, but we assume that $\phi=0$ just preceding the entangling interaction in each procedure, which corresponds to a shift in reference frame. Receiver atoms are described in the $\{|\text{c}\rangle,|\text{u}\rangle\}$ basis explained in \cref{sec:ExF}. 

All three protocols are \textit{heralded}, meaning that they conclude with a projective measurement, involving a destructive photon detection, the result of which indicates whether or not the protocol succeeded. This validation, which may be classically communicated back to the source nodes, is particularly important for probabilistic protocols since some measurement results indicate that the system has collapsed to an unusable state. Heralding is also important in lossy systems where messenger photons may be lost during photon collection, transmission through the communication channel, due to the protocol efficiency, or due to finite detection efficiency at the photon detector.

\subsection{Type-II (Two Photon) Protocol}
A cavity-enhanced variant of the two-photon protocol implemented in \cite{moehring2007entanglement} serves as a baseline. These networks are laid out symmetrically with two identical emitter cavities at the remote nodes. Each cavity couples light into fiber, and the two fibers meet at a central 50:50 \ac{npbs}. Following a synchronized excitation of the remote atoms, emitted photons overlap in space and time at the \ac{npbs}, concealing the particular source of each photon. A subsequent measurement of both photons in the computational basis, after the \ac{npbs}, projects the pair in the joint basis $\{|\text{e,e}\rangle, |\text{l,l}\rangle, |\Psi^+\rangle, |\Psi^-\rangle\}$ where $|\Psi^\pm\rangle \equiv \frac{1}{\sqrt{2}}\big(|\text{e,l}\rangle \pm |\text{l,e}\rangle\big)$ are the odd parity Bell-states. While the photon pair is destroyed by measurement, the correlation persists in the source atoms. 

Measurement of two photons separated by about $\Delta t$, corresponding to the $|\text{e}\rangle$ and $|\text{l}\rangle$ bins of the overlapped wavefunction, projects the photonic state to $|\Psi^\pm\rangle$ and heralds maximal-entanglement at the remote atoms (Note, with the other modalities, arrivals will be roughly coincident and the detectors will instead need to verify occupation of different polarization or frequency states). The protocol fails if fewer than two photons arrive or if a separable state is measured. An atomic excitation attempt from either emitter results in a photon detection with probability $P\approx P_{ex}\cdot P_1\cdot P_{L/2} \cdot P_{det}$, decomposed here as independent probabilities of faithful atomic excited state preparation $P_{ex}$, successful photon collection $P_1$, loss-free transmission $P_{L/2}$, and detector efficiency $P_{det}$. The four joint-photon states are equally likely, so $1/2$ of two-photon measurements herald success (protocol efficiency). Assuming identical $P$ for both sources, the overall per attempt success probability is $\mathcal{P}_2 = \frac{1}{2}P^2$.

Failure to eliminate which-path information (i.e. any information which could reveal the particular source of each photon exiting the \ac{npbs}) reduces fidelity of the resulting entangled states. An uneven \ac{npbs} ratio provides partial information about the path when photons exit different ports, reducing the fidelity of $|\Psi^-\rangle$ heralds, though not $|\Psi^+\rangle$. 
Incomplete overlap of photonic wavepackets allows the environment to distinguish photon sources, leading to a reduction in heralded state fidelity given by $\mathcal{F}_2 = \frac{1}{2}(1 + \langle\varphi_1,\varphi_2\rangle^2)$, where $\langle\varphi_1,\varphi_2\rangle$ is the overlap  between the two photon wavepackets integrated over space and time. Complete distinguishability in any dimension reduces the state to a classical mixture with residual fidelity of $1/2$. Infidelities arising from imperfect spatial overlap of the modes at the \ac{npbs} can be recovered by filtering output beams through a single mode fiber at the expense of reduced rates. Aligning temporal wavepackets requires routine calibration to ensure sub-$\unit{\nano\second}$ synchronization of emitters. The decaying sinusoidal form of emitted photon wavepackets are parameterized by the cavity $(g,\kappa,\gamma)$, so differences in the geometry, coating quality, and alignment of the two emitters create temporal mismatches that cannot be calibrated away. High-fidelity implementations require \textit{consistent} fabrication, but not necessarily low-loss or small volume mirrors. Cavities with low $P_1$ limit rate but not necessarily fidelity.

\subsection{Protocols with Strongly Coupled Recievers}
We consider two alternative RE protocols based on \ac{dit} and CPF effects, mediated by the transfer of a single photon. Unlike Type-II, these schemes are inherently asymmetric; the two atoms serve distinct roles and the photons only flow in one direction. In both cases, a cavity-coupled emitter atom (states sub-scripted $e$) generates a maximally-entangled photon ($p$) which gets directed towards a receiver atom ($r$) within another optical cavity. Strongly coupled to the cavity field, the receiver atom entangles with the arriving photon (and the emitter by proxy) before re-emitting the light. A beamsplitter transform re-casts the photon in the X-basis $|\pm\rangle_p = \big(|\text{e}\rangle_p \pm |\text{l}\rangle_p\big)/\sqrt{2}$ which may be observed without collapsing the atomic superposition. Detection of the photon verifies successful completion of the protocol without photon loss, heralding RE of the atoms. 

The emitter operates identically to nodes in a type-II scheme, exciting and out-coupling light with $P_{ex}\cdot P_1$. Propagation losses accrue over the full separation distance $L$, successfully transmitting light with $P_L$ (assuming losses are exponential with distance, $P_L\approx (P_{L/2})^2)$. Factoring out the influence of the receiver cavity, excitation attempts result in a detector click with end-to-end success probability $P'=P_{ex}\cdot P_1\cdot P_L \cdot P_{det}$.

The receiver interactions, based on \ac{dit} and CPF effects, are imperfect. Cavities with finite cooperativities leak and distort photons, adding an additional loss channel which limits success rates and introducing infidelity despite post-selection on heralding. We use \cref{eqn:dit} and \cref{eqn:cpf} to provide approximate formulae to estimate protocol efficiency and fidelity based on cavity quality (quantified by $C$ and/or $\mathcal{L}_B$). These formulae are approximate in the sense that finite-time photons necessarily have a non-$\delta$-function frequency spectrum, required by the Fourier limit, implying some noncompliance to the condition $\Delta_a=\Delta_c=0$. To reach an analytic form, we substitute near-resonant cavity reflectance (transmittance) with on-resonance values $r_{u/c}(\Delta)\approx r_{u/c}^\circ$ ($t_{u/c}(\Delta)\approx t_{u/c}^\circ$), in effect assuming that the spectral width of the photon around the carrier is narrow compared to receiver $g$ and $\kappa$. We call this condition \textit{perfect resonance} and will revisit the validity when we better understand the design constraints of receiver cavities.

\subsubsection{\Ac{dit} Protocol}
In our first scheme, the receiver cavity implements state-dependent filtration based on \ac{dit}, rejecting (reflecting) photons which do not imply a particular correlation between the atoms. By post selecting on transmission events, we exclusively project states where the emitter and receiver atoms occupy opposite qubit states.

To create the filter, the receiver atom is initialized to $|+\rangle_r=(|\text{u}\rangle_r + |\text{c}\rangle_r)/\sqrt{2}$ inside a balanced transmission cavity $\kappa_L=\kappa_R$. Receiver atoms in $|\text{c}\rangle_r$ suppress transmission of resonant photons through the cavity.  Between the arrival of the early and late photonic time-bins, a $\pi$-pulse inverts the receiver state $|\text{u}\rangle_r\leftrightarrow|\text{c}\rangle_r$ creating a time-dependent transmission window. Emitters which populate the early (late) time-bin occupy the state $|\text{u}\rangle_e$ ($|\text{c}\rangle_e$) after the photon generation sequence. The early (late) window is opened by receivers which occupy $|\text{u}\rangle_r$ before (after) the receiver $\pi$-pulse, so photons transmit freely for atoms in the joint state $|\text{uc}\rangle_{er}$ ($|\text{cu}\rangle_{er}$). In other words, emission bins and receiver windows align for odd-parity atomic states $\{|\text{uc}\rangle_{er}, |\text{cu}\rangle_{er}\}$. By projecting the photon in the $|\pm\rangle_p$ basis, the atom pair is projected into an entangled state. 

With perfect discrimination, postselection on transmission events \textit{carves} out the odd-parity subspace \cite{welte2017cavity}. Even and odd-parity atomic states are equally likely, so carving introduces a 50\% intrinsic protocol efficiency. In principle, we can also collect reflected photons which herald the even-parity subspace. By simultaneously monitoring reflections and transmissions, we could \textit{partition} the subspace rather than carving it, eliminating the intrinsic inefficiency at the cost of increased complexity. We limit attention to the transmission-only case for simplicity.

With finite-cooperativity cavities and non-transmissive losses, discrimination by \ac{dit} is imperfect. We can approximate the impact by tracking state evolution in the resonant-photon limit. Following transmission, the joint photon-emitter-receiver state (showing subscripts only on left-hand side) is 

\begin{equation}
    |\psi\rangle_{p e r} = \frac{\sqrt{P'}}{2}\Big(t^\circ_u|\text{ecu}\rangle + t^\circ_c|\text{ecc}\rangle + t^\circ_c|\text{luu}\rangle + t^\circ_u|\text{lcu}\rangle \Big),
\end{equation}

\noindent where we have discarded undetectable states from the wavefunction, resulting in sub-unity magnitude. Subsequent measurement of the photon in $|\pm\rangle_p$ projects the atomic state to

\begin{equation}\label{eqn:dit_pre-meas}
    |\widetilde{\Psi}^\pm\rangle_{er} \propto t^\circ_u |\Psi^\pm\rangle + t^\circ_c|\Phi^\pm\rangle,
\end{equation}

\noindent where the sign is determined by the result of the measurement. Leakage through the coupled-state cavity $|t_c^\circ|>0$ mixes even-parity states into the result. The fidelity of $|\widetilde{\Psi}^\pm\rangle$ to the ideal result $|\Psi^\pm\rangle$ is given by $\mathcal{F}^\circ_{DIT}= |t_u^\circ|^2 / (|t_c^\circ|^2 + |t_u^\circ|^2)$. This may be written as a pure, monotonic function of cooperativity

\begin{equation}\label{eqn:tfid}
    \mathcal{F}^\circ_{DIT} = \frac{(1+C)^2}{(1+C)^2 + 1} .
\end{equation}

\begin{figure*}
    \centering
    \includegraphics[width=\linewidth]{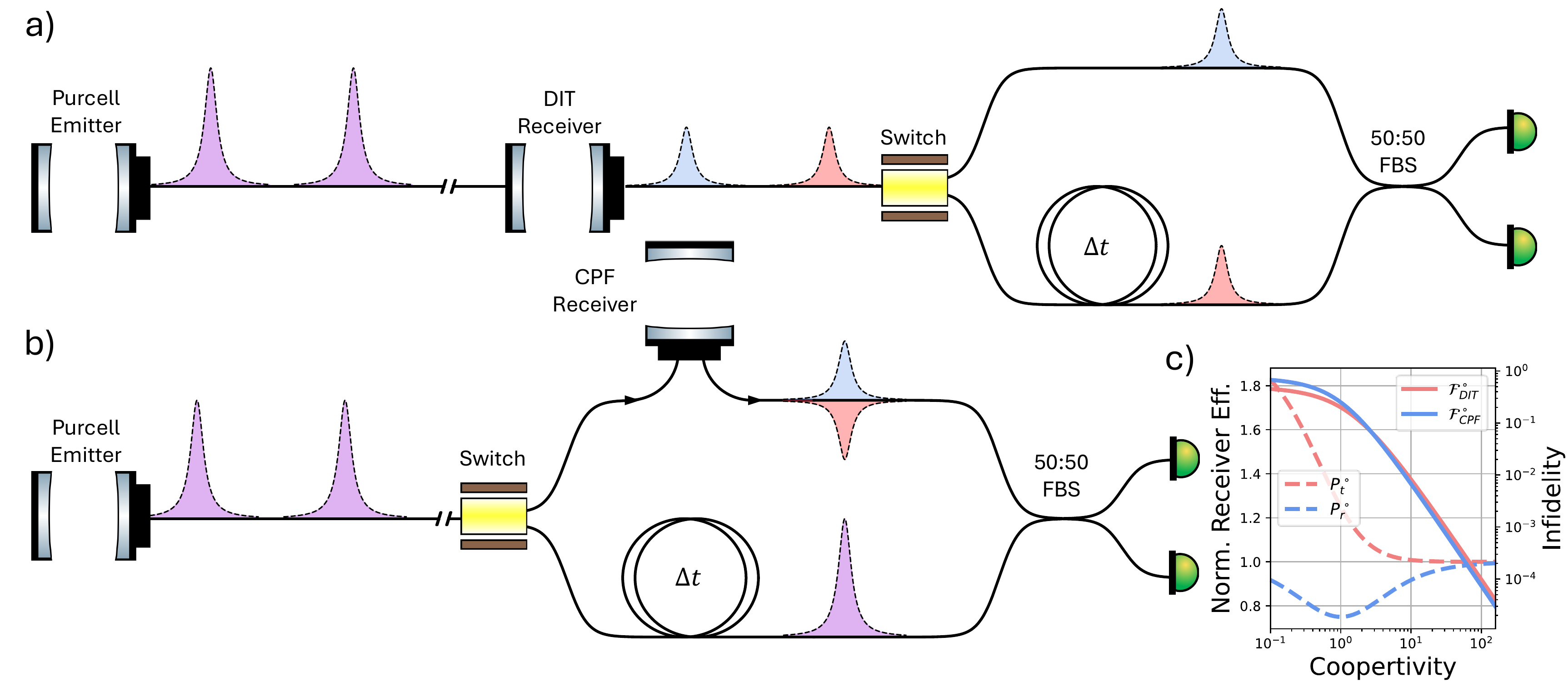}
    \caption{(a) \Acl{dit} layout: A Purcell cavity emits a time-binned photon and a \ac{dit}-type receiver rejects photons corresponding to even-parity atoms. The fiber beamsplitter (FBS) implements an $X$-basis measurement of transmitted photons, verifying atomic correlation (or a failed attempt) without collapsing atomic superposition. (b) \Acl{cpf} layout: an optical switch directs only late photons to a CPF receiver which applies an atomic state-dependent phase. Only $|\text{u,l}\rangle$ atom-photon states acquire this phase shift, $|\text{c,l}\rangle$, $|\text{u,e}\rangle$, $|\text{c,e}\rangle$ do not, leading to a controlled-Z like entangling interaction. (c) $P_t$ ($P_r$) and infidelity of the \ac{dit} (\ac{cpf}) roles as a function of cooperativity, assuming $\kappa_L\simeq \kappa$. }
    \label{fig:layouts}
\end{figure*} 

The total success probability per-attempt is  $\mathcal{P}_{DIT}^\circ=\frac{1}{2}P\cdot P_t^\circ$ where $P_t^\circ\equiv |t_c^\circ|^2 + |t_u^\circ|^2$ is the probability of \textit{any} transmission normalized to the odd-parity population before transmission. Defined this way, $P_t^\circ$ is unity for lossless, infinite-cooperativity cavities and the factor of $1/2$ encodes the protocol efficiency. Rewriting, we get

\begin{equation}\label{eqn:tp}
    P_t^\circ = \Big(\frac{2 \kappa_L}{\kappa}\Big)^2 \Big[1 + \frac{1}{(1+C)^2}\Big].
\end{equation}

Low cooperativity impairs cavity discrimination, boosting transmission probability with false-heralding events that degrade fidelity, see \cref{fig:layouts}c. If bad-losses dominate transmission, the ratio $2\kappa_L/\kappa$ also limits success probability. 

Misaligned input modes will strongly reflect from the input mirror instead of interacting with the internal cavity mode. A mode-matching parameter $\xi$, defined as the mode overlap integral between the input and cavity modes, reduces intensity transmission to $|\widetilde{t}(\omega)|^2 =\xi |t(\omega)|^2$ and increases reflected intensity to $|\widetilde{r}(\omega)|^2 = (1-\xi) + \xi |r(\omega)|^2$. Imperfect mode-matching decreases coupled and uncoupled transmission in proportion, maintaining $\mathcal{F}^\circ_{DIT}$ but reducing $\mathcal{P}^\circ_{DIT}$ by a factor of $\xi$ for transmissive projections. Note that for equivalent protocols which herald on reflections, input misalignment increases reflected intensity from uncoupled cavities (ideally $r_u=0$) by a far greater factor than the increase for coupled cavities, amplifying infidelity. 

Rejection of unwanted photons is most effective when $\Delta_a=\Delta_c=0$ and weakens with photon detuning. The coupled-cavity resonances, split $\pm g$ from the bare-cavity resonance, will leak photons detuned by $\pm g$. Unwanted light in the pre-measurement state (\cref{eqn:dit_pre-meas}) degrades fidelity. In order for the perfect resonance approximation to remain valid, photon spectral widths $\sigma_\omega$ about the carrier must remain well inside of those coupled-cavity resonances, requiring $\sigma_\omega\ll g$.

\subsubsection{\ac{cpf} Protocol}
In this scheme, originally proposed by Duan and Kimble, photonic qubits reflect from a receiver apparatus with a phase which depends on the receiver-photon state based on the \ac{cpf} effect\cite{duan2004scalable}. Local rotations recast this controlled-Z interaction into a CNOT gate which extends the maximally entangled emitter-photon state into a GHZ state including the receiver as well. Readout of the photon in $|\pm\rangle_p$ heralds completion of the protocol without collapsing the atomic Bell state. The interaction does not carve the detectable state-space and all detection events herald success and the protocol efficiency is unity.

The receiver apparatus consists of an imbalanced cavity preceded by a photon-state dependent pre-filter, see \cref{fig:layouts}b. The receiver atom is initialized to $|+\rangle$ inside the cavity so that all photon-receiver computational basis states are equally likely. The pre-filter converts photons to dual-rail, directing the $|\text{e}\rangle$ photons to a $\Delta t$ delay line and $|\text{l}\rangle$ photons into the \ac{cpf} cavity input mode. The $|\text{l}\rangle$ photons reflect with $r_c$ or $r_u$, depending on the receiver state. After the cavity interaction, the early and late paths overlap at a \ac{npbs}, concealing which-path information, prior to an $X$-basis readout of the photon. Ideally, $\angle r_u = \pi$ is the relative phase shift of $|\text{l,u}\rangle$ relative to the other basis states.

We assume a lossless delay path and estimate cavity reflectances in the perfectly-resonant limit. After cavity and delay paths overlap at the \ac{npbs}, the system state is

\begin{equation}
    |\psi\rangle_{p e r} = \frac{\sqrt{P'}}{2}\Big(|\text{ecc}\rangle\ + |\text{ecu}\rangle\ + r^\circ_c|\text{luc}\rangle\ + r^\circ_u|\text{luu}\rangle\Big),
\end{equation}

\noindent again discarding undetectable population from the wavefunction. A Hadamard gate on the receiver atom transforms $|\text{c}\rangle_r\mapsto\frac{1}{\sqrt{2}}\big(|\text{c}\rangle + |\text{u}\rangle \big)$ and $|\text{u}\rangle_r\mapsto\frac{1}{\sqrt{2}}\big(|\text{c}\rangle - |\text{u}\rangle \big)$. Then, subsequent $X$-basis measurement of the photon kicks back a phase and projects the final atomic state

\begin{equation}\label{eqn:cpf_pre_meas}
    |\widetilde{\Phi}^\pm\rangle_{er} \propto \Big(|\text{uu}\rangle \pm \frac{r_c^\circ + r_u^\circ}{2}|\text{cu}\rangle \pm \frac{r_c^\circ - r_u^\circ}{2}|\text{cc}\rangle\Big),
\end{equation}

\noindent converging to $|\Phi^{\pm}\rangle$ for ideal reflectances $r_c^\circ = -r_u^\circ = 1$. Fidelity of $|\widetilde{\Phi}^\pm\rangle$ to the $|\Phi^\pm\rangle$ state is given by 

\begin{equation}\label{eqn:rfid}
    \mathcal{F}_{CPF}^\circ = \frac{|1 - \frac{1}{2}(r_u^\circ-r_c^\circ)|^2}{2 + |r_u^\circ|^2 + |r_c^\circ|^2}.
\end{equation}

\noindent Each attempt heralds success with probability $\mathcal{P}_{CPF}^\circ = P\cdot P_r^\circ$ where $P_r^\circ\equiv \frac{1}{4}\big(2 + |r_u^\circ|^2 + |r_c^\circ|^2\big)$ is the intensity reflected from the receiver apparatus averaged over all four basis states. The magnitudes of $|r_u|$ and $|r_c|$ are limited by bad losses, and $|r_c^\circ|$ is further reduced by finite cooperativity. Absent any bad losses $\kappa_L/\kappa=1$, $\mathcal{F}$ increases monotonically with $C$ and $P_r$ also increases with $C$ when $C>1$ (\cref{fig:layouts}c). Accounting for these losses, transmission may be increased to outpace bad loss, at the cost of reduced cooperativity, presenting a tradeoff. 

Misalignment between the incident photon mode and cavity causes some fraction of light to reflect promptly with $r_p\equiv 1$, regardless of the receiver state. Light coupling back into the fiber will be a mixture of photons which underwent the desired \ac{cpf} and these prompt reflections which did not. To understand the nature of this problem, we imagine a toy-model where $r_u^\circ$ is replaced by a mixed coefficient $\widetilde{r}_u^\circ = w_1\cdot r_u^\circ + w_2\cdot r_p$ with real-valued weights $w_{1,2}$ subject to $|w_1|^2+|w_2|^2=1$. In this case, $\widetilde{r}_u^\circ > r_u^\circ$ implying incomplete interference in \cref{eqn:cpf_pre_meas} and a corresponding loss of fidelity. Realistically, mode-mismatch can lead to complicated errors non captured by this simple model, but the challenge of mixed-coefficients remains.  Recall that with \ac{dit} we opted for the transmissive variant of the scheme, rather than reflective, to eliminate mode-matching based infidelity at the cost of photon loss. Due to the asymmetry of the \ac{cpf} cavity, there is no equivalent transmissive version of the \ac{cpf} protocol, making this mode-matching problem a major practical challenge. Atomic systems often require a fiber-to-free-space interface at the cavity and relying on perfect mode-matching here is not practical. Potentially, solid-state systems with very tight mode volumes coupled to only one waveguide mode (such as in photonic crystal cavity-waveguide system) might circumvent some of these challenges~\cite{YaoPRB2009,NguyenPRL2019}.

The phase-flip interaction is also band-limited. The relative phase shift $\phi = \angle r_u(\omega)-\angle r_c(\omega)$ is exactly $\pi$ on resonance, but this breaks down when the photon has a finite spectral width $\sigma_\omega$. These frequency-dependent phase-shifts distort temporal wavepackets, leading to incomplete interference between coupled and uncoupled reflections as well as which-bin information after combination at the beam-splitter. Near resonance, the relative phase shift varies like the bare-cavity Lorentzian phase $\phi\approx \arctan(-\Delta/\kappa)$, requiring $\sigma_\omega\ll\kappa$ for the perfect resonance approximation to remain valid.

\section{Results} \label{sec:results}
In all three protocols highlighted, uncollectable cavity decay $\kappa_B>0$ or finite coupling strength $g$ limit success rate and/or fidelity of the resulting state. A realistic mirror fabrication process will introduce some bad losses $\mathcal{L}_B$ and set a lower limit on $R$, so non-ideal performance cannot be avoided. Here we compare performance of the three protocols when all protocols are subject to the same fabrication process limitations. 

In what follows, we provide specific cavity constructions, efficiency comparisons to understand tradeoffs intrinsic to the protocol, and finally rate comparisons for realistic experimental conditions. To limit scope, we only consider limitations intrinsic to the protocol and the quality of the cavity. Of note, we unrealistically assume atomic excitation, photon propagation, and photon detection all proceed with unit efficiency $P_{ex},\, P_{L(/2)},\, P_{det}=1$. We also assume perfect mode-matching $\xi=1$ at interfaces to receiver cavities. We conclude with a brief discussion of how these various inefficiencies could impact our results. 

\subsection{Constructions}
We assume full freedom to determine left and right mirror transmission and select any cavity length within the bounds of cavity stability $\ell < R$ to compensate for imperfections in mirror fabrication parameterized by $(\mathcal{L}_B, R)$. For concreteness, we propose a definition of optimality and use this definition to reduce the design space to a unique construction for each cavity role (emission, selective transmission, or \ac{cpf} reflection).

\subsubsection{Emission Cavities}
Emission cavities are constructed to maximize out-coupling efficiency $P_1$. We claim that the emission cavity does not inherently limit heralding fidelity for any of the protocols for the following reasons. For type-II protocols, cavity $g,\kappa, \gamma$ do not restrict heralded fidelity as long as the temporal profile of the photon wavepackets are identical between the two sources. Fidelity degrades in the \ac{dit} and \ac{cpf} protocols when spectral widths of the photons $\sigma_\omega$ are too broad for uniform spectral response at the receiver, but we cast this as a limitation of the receiver since, in principle, a fast enough receiver could interact with any emitted photon faithfully. Furthermore, we could consider elongated-excitation pulses (rather than pseudo-instantaneous population transfer), pulse-modulation techniques \cite{lukens2016frequency} or cavity filtration to narrow emitted photon $\sigma_\omega$. In either case, the chief purpose of the cavity is collecting the photon, not shaping it. 

Selection of mirror transmission allows tuning of $\kappa_L$ without impacting cavity $g$ or $\kappa_B$. Optimizing over $\kappa_L$, we reduce the search space for the maximum of $P_1$ to the contour 

\begin{equation}
    \kappa_L^\star(\ell) = \sqrt{g^2 + \kappa_B(\gamma + \frac{g^2}{\gamma} + \kappa_B)} ,
\end{equation}

\noindent where the $\star$ indicates optimality. When $\kappa_B\rightarrow 0$ this simplifies to the \textit{optimal cavity regime} $\kappa_L = g$ identified in \cite{cui2005quantum}. Inspecting the factors of $P_1$ (\cref{eqn:p1}), $\eta_c$ is maximized when $C$ is maximized at length $\ell_o$ (\cref{sec:purcell}), but $\eta_{ex}$ asymptotically approaches max efficiency when $\ell\rightarrow 0$ as $\kappa_L\gg\gamma$. We use a numerical line-search of resonant lengths less than $\ell_o$ to identify the optimal cavity length $\ell^\star$, since different length scalings in $g$ and $\kappa$ obscure a closed form result. 

\subsubsection{Transmission Cavities}
In \ac{dit} protocols, properties of the transmission cavity affect both $\mathcal{F}_{DIT}^\circ$ and $P_t^\circ$, but neither quantity can be used directly as a reward function to optimize cavity parameters. Fidelity is monotonic in $C$, but naively reducing mirror transmission to boost $C$ also suppresses transmission rates. Likewise, $P_t$ reaches a maximum when $C=0$ because false-heralds boosts ``success" rates despite yielding unusable classical mixtures rather than Bell states. Instead, we specify a minimum acceptable fidelity $\mathcal{F}_{min}$, implying a minimum cooperativity $C_{min}\equiv \sqrt{\mathcal{F}_{min}/(1-\mathcal{F}_{min})}-1$, derived from \cref{eqn:tfid}, and maximize $P_t$ within the in-spec region. The optimum always saturates this minimum specification, reducing the design space to a single curve

\begin{equation}
    \kappa_L^\star(\ell) = \kappa_R^\star(\ell) = \frac{1}{2}\Big(\frac{g^2}{C_{min}\gamma} - \kappa_B\Big)  ,
\end{equation}

\noindent where transmission is adjusted to compensate changes in cavity length, holding $C$ constant at $C_{min}$. Restricted to this line, \cref{eqn:tp} implies $\sqrt{P_t}\propto \kappa_L/\kappa = 1-C_{min}/C_o$, where $C_o\equiv g^2/\kappa_B\gamma$ is the zero-transmission cooperativity. $P_t$ reaches a maximum at $\ell^\star = \ell_o$ where $C_o$ is largest.

\subsubsection{Reflection Cavities}
In the \ac{cpf} protocol, $\mathcal{F}_{CPF}^\circ$ and $P_r$ both reach a maximum when $C(2+C) = C_o$, most easily shown by again replacing ratios of $\kappa_i$ with ratios of $C$ and $C_o$ in \cref{eqn:rfid} and differentiating with respect to $C$. By no coincidence, this is precisely the condition that $r_c^\circ = |r_u^\circ| = 1 - \frac{2}{2+C}$, allowing for complete destructive interference of the $|\text{cu}\rangle$ term in the heralded state. Since transmission may be used to tune $C$ without impacting $C_o$, we treat this condition as a constraint on transmission (defining $\kappa_L^\star(\ell))$, confining our search to a 1D contour parameterized by $\ell$. On this line,

\begin{equation}
\mathcal{F}^\circ_{DIT}\Big|_{\kappa_L^\star} = \frac{1+C_o}{2+C_o}\quad P_r^\circ\Big|_{\kappa_L^\star} = \frac{1+(1+C)^2}{(2+C)^2},
\end{equation}

\noindent which are simultaneously maximal at $\ell^\star = \ell_o$. Since $\max_\ell(C)$ is independant of $R$, the process $\mathcal{L}_B$ exclusively determines fidelity.

\subsection{Success Probability Comparisons}
These cavity constructions above specify a $(g,\kappa_i, \gamma), i\in\{L,R,B\}$ for each type of cavity across the full space of fabrication processes. With our assumption that  $P_{ex}=P_{L(/2)} = P_{det} = 1$, failed collection, protocol indeterminism, and imperfect receivers are the only loss channels. Per attempt success probabilities are fully determined at each $(R, \mathcal{L}_B)$ for each scheme. 

Due to its prominence in atomic platforms, we take type-II entanglement as a baseline and plot the success probability advantage $(\mathcal{P}_i-\mathcal{P}_2)/\mathcal{P}_2, i\in\{\text{DIT}, \text{CPF}\}$ of converting to a \ac{scr} protocol but holding the type of photonic qubit constant. For brevity, we only show the time-bin comparison, but the tradeoff looks similar for the other modalities. In the context considered, type-II fidelity is perfect, so this conversion necessarily introduces some infidelity, which we quantify. 

\subsubsection{\ac{dit} Probability Advantage}
\begin{figure}
    \centering
    \includegraphics[width=\linewidth]{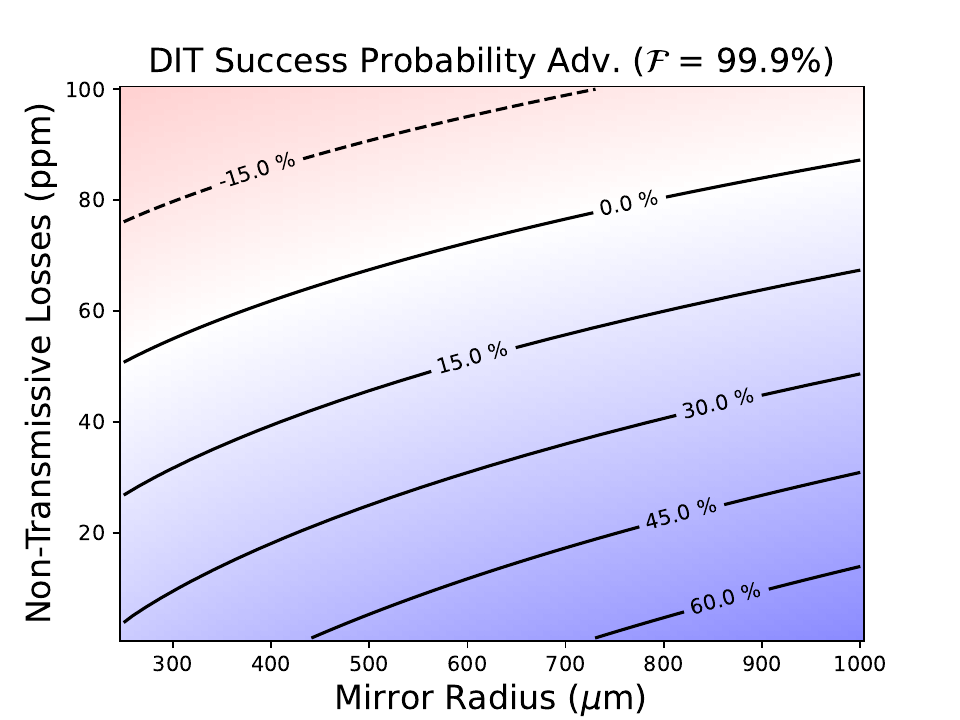}
    \caption{Per-attempt success probability advantage (not rate) from converting entanglement protocol from type-II to \ac{dit}, modeled across a range of constraints on size $R$ and non-transmissive loss $\mathcal{L}_B$ which limit both emission and \ac{dit} cavities.}
    \label{TVD_adv}
\end{figure}

Receiver cavities for the \ac{dit} protocol are optimized for a specific target fidelity, set here to $\mathcal{F}_{min} = 1-10^{-3}$, high enough not to be a limiting factor in any contemporary RE demonstrations, but low enough not to restrict success rates needlessly. Figure~\ref{TVD_adv} compares the success probability advantage of the \ac{dit} protocol compared to the type-II protocol.

Both \ac{dit} and type-II protocols feature an intrinsic inefficiency of $\frac{1}{2}$ and the cavity collection of at least one photon with $P_1$. The remaining difference arises from the relative strength of $P_T$ over $P_1$. Over the range considered, bad losses have the strongest impact on relative rate performance. Transmission based protocols become advantageous roughly when $\mathcal{L}_B\lesssim60$ \ac{ppm} for this target fidelity, where transmission is more reliable than collection of a second photon from an emitter cavity. Collection improves with decreased $R$, making the $\mathcal{L}_B$ required for breakeven \ac{cpf} performance with the \ac{dit} protocol increasingly stringent.
Tighter fidelity specifications (and larger $C$) require decreased receiver transmission, further reducing tolerance to bad loss, and vise versa. For example, $\mathcal{F}_{min}=1-10^{-4}$ requires $\mathcal{L}_B\lesssim20\text{ppm}$, but $\mathcal{F}_{min}=1-2\times 10^{-3}$ permits $\mathcal{L}_B\gtrsim 100\text{ppm}$.

\subsubsection{\ac{cpf} Probability Advantage}
\begin{figure}
    \centering
    \includegraphics[width=\linewidth]{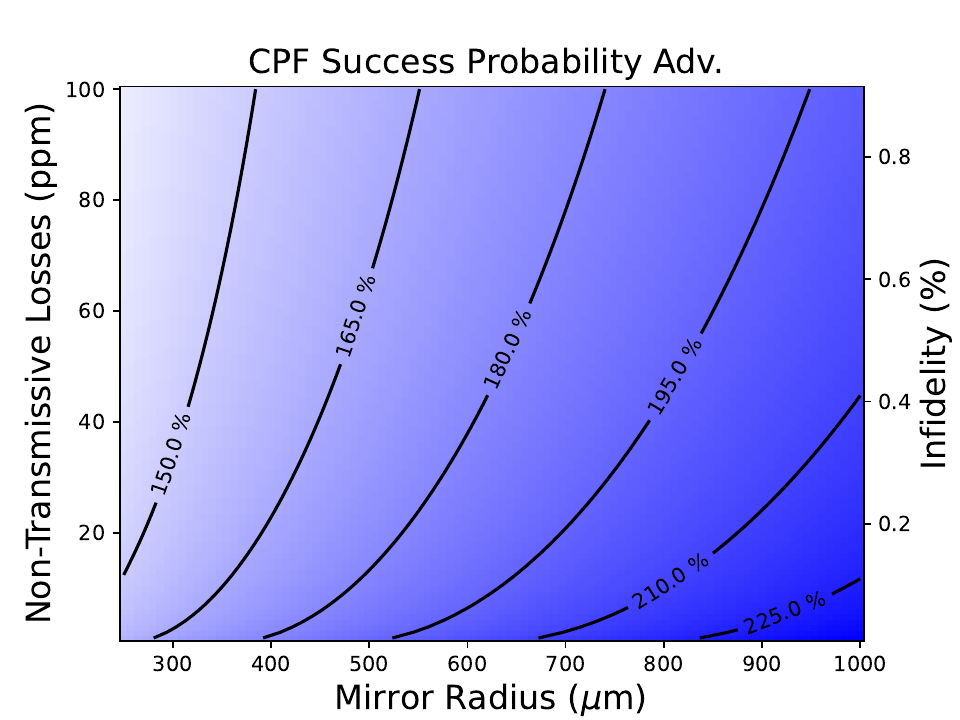}
    \caption{Per-attempt success probability advantage from converting entanglement protocol from type-II to \ac{cpf}, modeled over a range of constraints on minimum achievable mirror curvature $R$ and non-transmissive loss $\mathcal{L}_B$ which limit both emission and \ac{cpf} cavities. Since $\mathcal{L}_B$ also determines $\mathcal{F}_{CPF}^\circ$ with our construction, we provide \ac{cpf} infidelity as an alternative y-axis.}
    \label{fig:CPF_adv}
\end{figure}

In the \ac{cpf} protocol, the fidelity of the protocol is heavily dependent on the quality of mode-matching between the incoming beam and the cavity mode. Perfect mode-matching is quite impractical in the cavity scheme considered in this paper, described in Figure~\ref{fig:micro_cav}. In this section, we nevertheless make the highly optimistic assumption of perfect mode-matching to analyze the potential benefit of the \ac{cpf} protocol.  Under this assumption, the infidelity is entirely dictated by bad losses, and the one-to-one correspondence between $\mathcal{L}_B$ and the infidelity $\mathcal{F}^\circ_{CPF}$ is provided as a secondary $y$-axis in \cref{fig:CPF_adv}. 

The \ac{cpf} protocol has unit protocol efficiency so an efficiency advantage can be observed if $P_R > \frac{1}{2}P_1$, comfortably achievable over the full range of processes surveyed. At small $R$, high achievable $P_1$ slightly reduces the advantage, but the \ac{cpf} protocol still enjoys more than double the efficiency even at $R\approx250\unit{\micro\meter}$. Fidelity is the clear limitation of this protocol. The free-parameters do not allow the tradeoff of rate for improved fidelities, so performance is beholden to $\mathcal{L}_B$. The same $\mathcal{F}=99.9\%$ target laid out in the previous section requires $\mathcal{L}_B\lesssim 10\text{ppm}$. But, for processes which permit these low-loss mirrors, \ac{cpf} is easily the most efficient protocol considered.

\subsection{Rate Examples}
Per-attempt success probability $\mathcal{P}$ is related to success rate as $R_{suc} = \mathcal{P}\cdot R_{att}$ where $R_{att}$ is the attempt rate. Success probability is an incomplete indicator for $R_{suc}$ because $R_{att}$ varies due to several extrinsic factors. Here we consider two hypothetical experiments to explore how timing details impact rate. The first experiment $\mathcal{E}_1$ is modeled after contemporary RE demonstrations where attempts are made between two nodes in the same lab. $\mathcal{E}_2$ envisions a more ambitious long-haul system multiplexed to allow several entanglement attempts to proceed in parallel \cite{monroe2013scaling}. Both experiments are modeled with specific transitions in Barium ions, outlined in \cref{apx:ba_trans}. These examples are by no means a complete representation of the diversity of potential designs, but do provide the opportunity to highlight several protocol limitations.

We determine attempt rate as $R_{att}=\tau^{-1}$ where $\tau$ is the total attempt cycle time. A basic cycle might proceed as follows: After some electronic latency $t_E$ as control signals propagate to laser modulators, atoms are optically pumped for a duration $t_P$. Time-binned qubits also require a preparatory $\pi/2$-pulse, lasting $\frac{1}{2}t_\pi$. The photon collection window, beginning in sync with the excitation pulse(s), must accommodate the photon length. This lasts for a single bin-width $s$ for polarization and frequency qubits and for two bins separated by a $\pi$-pulse for time-binned qubits and $2\cdot s + t_{\pi}$. Outbound photonic qubits propagate to their targets in $t_{tx}$ and the classical measurement results are relayed back to the sources after $t_{rx}$. The meet-in-the-middle architecture of type-II protocols can halve propagation time for the outbound photon $t_{tx}$ and inbound classical verification signal $t_{rx}$, offering a significant time-saving if propagation distance $L$ is large.

Each one of these processes must be run serially for each entanglement attempt. In $\mathcal{E}_1$, each atom remains in the cavity during initialization and while awaiting the heralding signal, introducing dead-time while the atom is not interacting with the cavity mode. In $\mathcal{E}_2$, many pairs of atoms share the same communication line in an attempt to maximize fiber utilization. In particular, we envision a system where pre-initialized ions are shuttled into the cavity for emission/reception and cycled out to storage during propagation delays, freeing the cavity for subsequent entanglement attempts. This takes $t_E, t_P, t_{tx}$ and $t_{rx}$ offline, as well as the preparatory $\frac{1}{2}t_\pi$ for time-binned schemes, in exchange for the added shuttling time $t_S$, see \cref{tab:ctime_defs}. Shuttling is likely slower than most local operations but much faster than $t_{tx}$ and $t_{rx}$ if propagation distances are long, providing an opportunity for savings in long-haul systems. 

\begin{table}
\caption{\label{tab:ctime_defs}
Cycle time definitions used in $\mathcal{E}_1$ and $\mathcal{E}_2$ depending on whether the protocol is multiplexed (MUX) or implemented with time-binned qubits (TB).}
    \begin{ruledtabular}
        \begin{tabular}{ccc}
        \textrm{MUX}&
        \textrm{T.B.}&
        \textrm{Cycle Time}\\
        \colrule
        \xmark&  \xmark & $t_E + t_P + t_{tx} + t_{rx} + s$\\ 
        \xmark & \cmark & $t_E + t_P + t_{tx} + t_{rx} + 1.5\cdot t_{\pi} + 2\cdot s$ \\ 
        \cmark & \xmark & $t_S+ s$\\ 
        \cmark & \cmark & $t_S +  t_{\pi} + 2\cdot s$\\
        \end{tabular}
    \end{ruledtabular}
\end{table}

Emitters and receivers constrain the temporal shape of photons, and photon bin widths $s$ are chosen to accommodate a large portion of the wavepacket without inflating cycle time. A bin width of $s_o = N\cdot K^{-1}$, with number of collection lifetimes $N>1$, ensures that less than $e^{-N}$ of photon population falls outside of the window. For single photon protocols, we may need to elongate photons to ensure that spectral width does not exceed receiver bandwidth. If $s_o$ is insufficient, we set $s_{DIT} = S\cdot\pi g^{-1}_r$ or $s_{CPF} = S\cdot\pi\kappa^{-1}_r$ where $S\gg 1$ is a safety factor which constrains the photon spectrum to a fraction of the receiver bandwidth. Timing parameters and bin definitions are outlined in \cref{tab:combined}.

\begin{table}[h]
\caption{\label{tab:combined}Timing parameters and definitions for hypothetical RE experiments.}
    \begin{ruledtabular}
        \begin{tabular}{cccc}
        \textrm{ }&
        \textrm{Symbol}&
        \textrm{$\mathcal{E}_1$}&
        \textrm{$\mathcal{E}_2$}\\
        \colrule
        $\pi$-Time & $t_{\pi}$ & $1\unit{\micro\second}$ & $200\unit{\nano\second}$ \\
        Pump Time &  $t_{P}$ & $300\unit{\nano\second}$ & (offline)\\ 
        Propagation Time &  $t_{L(/2)}$ & $10\unit{\nano\second}$ & (offline)\\
        Electronic Latency & $t_{E}$ & $400\unit{\nano\second}$ & (offline) \\
        Shuttling & $t_S$ & 0 & $1\unit{\micro\second}$\\
        Collection Lifetimes & $N$ & \multicolumn{2}{c}{3}\\
        Band Safety Factor & $S$ &  \multicolumn{2}{c}{10} \\ \hline
        Minimum Bin Width & $s_o$ & \multicolumn{2}{c}{$N\cdot\Gamma^{-1}$} \\         \ac{dit} Bin Width & $s_{DIT}$ & \multicolumn{2}{c}{$\text{max}(s_o, S \pi g_r^{-1})$}\\ 
        \ac{cpf} Bin Width & $s_{CPF}$ & \multicolumn{2}{c}{$\text{max}(s_o, S\pi\kappa_r^{-1})$} \\ 

        \end{tabular}
    \end{ruledtabular}
\end{table}

\begin{figure}[h]
    \centering
    \includegraphics[width=\linewidth]{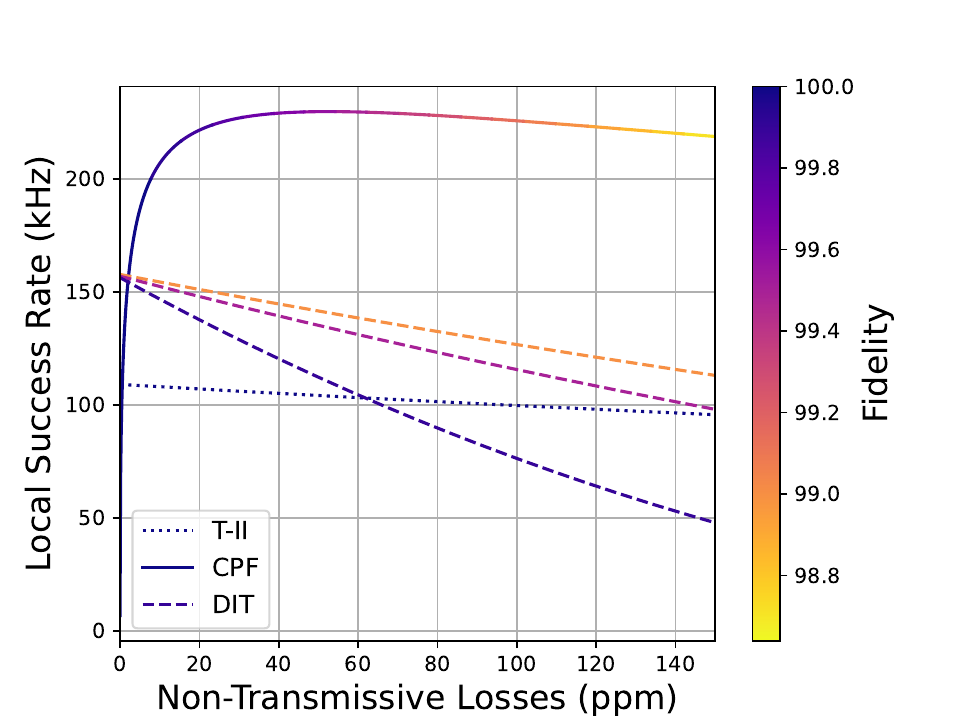}
    \caption{Estimated rates for three different protocols implemented with time-bin photons in a hypothetical near-term RE demonstration in $\text{Ba}^{+}$, described in \cref{tab:combined} column $\mathcal{E}_1$.  All cavities are implemented with $R=400\unit{\micro\meter}$ curved mirrors and modeled over a range of non-transmissive losses. Varying protocol fidelities are labeled on the color axis. \ac{dit} cavities are tuned fro $1-\mathcal{F}_{min}\in\{1, 0.5, 0.2\}\%$. \ac{cpf} cavities are always tuned for maximal fidelities, leading to a decline in bandwidth at low $\mathcal{L}_B$. }
    \label{fig:tb_rates}
\end{figure}

In \cref{fig:tb_rates} we model $R_{suc}$ for $\mathcal{E}_1$ across a range of $\mathcal{L}_B$ limitations at a fixed mirror curvature $R = 400\unit{\micro\meter}$. We plot rates for time-binned qubits only, highlighting advantages inherent to protocols rather than differences between qubit types. Similar patterns are observed for the other photon modalities. \ac{dit} cavities are tuned for three different target fidelities $1-\mathcal{F}_{min}\in\{1, 0.5, 0.2\}\%$, 

For relaxed fidelity standards (e.g $\mathcal{F}\approx99\%$), \ac{cpf} succeeds at the highest rate due to the doubled protocol efficiency. For loftier fidelity targets (e.g. $\mathcal{F}\gtrsim 99.9\%$), \ac{cpf} requires very low-loss and inherently low-bandwidth receivers, requiring a slow $R_{att}$ which diminishes the rate advantage. By contrast, type-II heralds at a modest rate but delivers fidelities not limited by the mirror imperfections we consider. If mirror fabrication abilities allow for $\mathcal{L}_B\lesssim 50\text{ppm}$, \ac{dit} provides an intriguing compromise, offering a meaningful rate improvement over type-II at the cost of a likely negligible contribution of infidelity.

\begin{figure}[h]
    \centering
    \includegraphics[width=\linewidth]{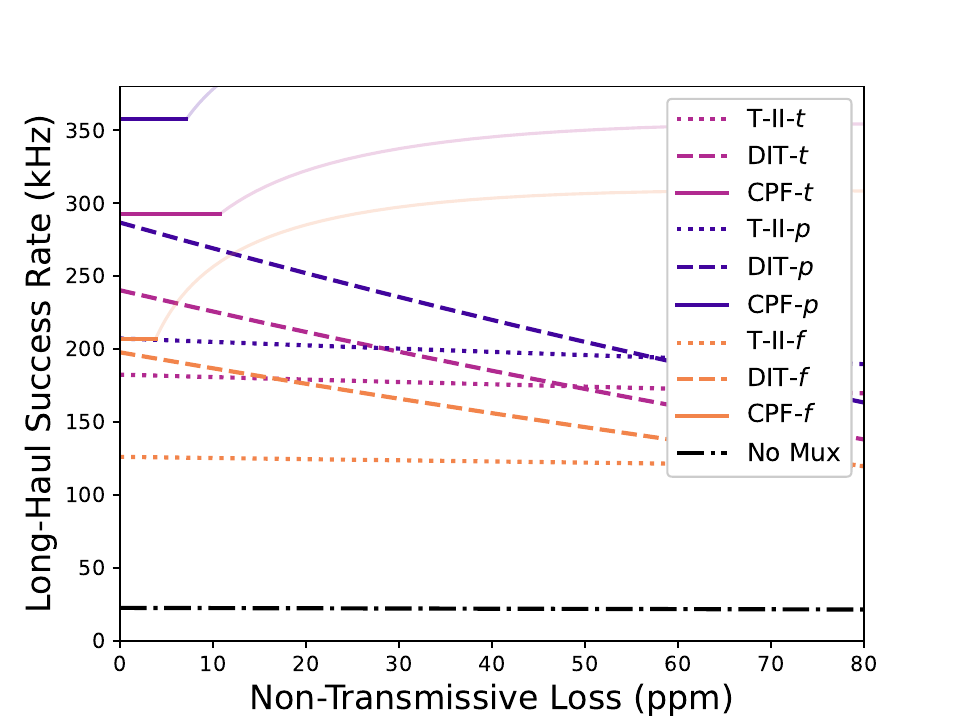}
    \caption{Three protocols implemented on three different types of photonic qubits on the hypothetical multiplexed system $\mathcal{E}_2$ specified in  \cref{tab:combined}. All cavities are implemented with $R=400\unit{\micro\meter}$ curved mirrors and modeled over a range of non-transmissive losses. A fidelity of $\mathcal{F}_{min}=99.9\%$ is targeted. Under-performing \ac{cpf} cavity configurations (in terms of $\mathcal{F}$) are displayed with lightened lines. Additional leakage pathways are introduced to over-achieving  \ac{cpf} cavities, in order to maintain bandwidth when $\mathcal{L}_B$ is low.  Also shown in black, the non-multiplexed $\mathcal{E}_1$ using polarization qubits with a type-II protocol with transmission delay $t_{tx}+t_{rx}\simeq 5\unit{\micro\second}$ corresponding to a $1\unit{\kilo\meter}$ separation.}
    \label{fig:all_rates}
\end{figure}

In \cref{fig:all_rates}, we model $R_{suc}$ of $\mathcal{E}_2$ for all three protocols implemented with all three photon modes. Again, we assume $R=400\unit{\micro\meter}$ and vary non-transmissive losses. We set a single target fidelity of $\mathcal{F}_{min}=99.9\%$. At low $\mathcal{L}_B$, \ac{dit} receivers  configured for maximum fidelity have a limited bandwidth which slows rates. Here, we maintain wider bandwidths when $\mathcal{F}^\circ_{CPF} < \mathcal{F}_{min}$ by purposefully adding cavity loss (via $\kappa_R$). A better solution might involve using bin-width $s$ as a third free-parameter to trade off fidelity for rate, but this requires a complete treatment of the full photon spectrum and deviates from the constructions outlined and corresponding formulae. 

With this high fidelity target, type-II protocols are the only reasonable choice for higher-loss mirrors. \Ac{dit} becomes potentially advantageous for $\mathcal{L}_B\lesssim60\text{ppm}$. \ac{cpf} protocols are only feasible at very low losses, although band limitations make rates comparable with \ac{dit}. This model neglects mode-matching errors, but recall that this leads to infidelity in \ac{cpf} protocols and merely a rate-reduction in \ac{dit}.

Polarization qubits benefit from strong $P_1$ (unlike frequency qubits) and short photons (unlike time-binned qubits), usually leading to the best rates for a particular choice of protocol. This model does not account for the significant challenges to transmitting polarization qubits over long distances with high fidelity discussed in \cref{sec:qbs}. Shuttling $t_S$ dominates the cycle time $\tau$, lowering the relative time cost for using time-bin qubits. Frequency qubits are consistently the worst performers due to low $P_1$ - a cost paid twice-over in the type-II scheme. We do not consider photon-modality specific sources of infidelity, but differing transition strengths do impact performance. In particular, the large $\mu$ of the transition driven by time-binned qubits loosens the $\mathcal{L}_B$ specification required to achieve $\mathcal{F}_{min}$ in the \ac{cpf} protocol, enabling higher bandwidth operation (compared to polarization and frequency qubits). 

Suppose we repeated analysis with inefficient $P_{ex},\, P_{det},\, P_{L(/2)},\,  \xi <1$. Since the type-II scheme require two separate excitation and detection events to succeed while both \ac{scr} protocols only require one each, we would expect these inefficiencies to slow type-II rates relative to the other protocols. While the type-II protocol also depends on the lossless transmission of two separate photons, these photons only need to propagate over half the node separation distance $L$. Given the exponential nature of propagation losses with distance, we expect roughly the same rate reduction factor $P_L\approx P_{L/2}^2$ for all protocols implemented with the same type of photon, effectively re-scaling the $y$-axis but maintaining the relative advantages between protocols. This layer of consideration would be more interesting when comparing transitions with different wavelength $\lambda$, since $\lambda$ strongly affects fiber loss rate. Finally, imperfect cavity mode matching leads to fidelity loss in \ac{cpf} but only rate reduction in type-II and \ac{dit}. The $\mathcal{L}_B$ requirements to achieve competitive fidelities with \ac{cpf} were already quite stringent, but the extra challenge of imperfect mode-matching could make practical implementation entirely infeasible. 
\section{\label{sec:conclude}Conclusion}
In this paper, we discussed several methods to entangle ions and photons as well as three protocol classes which use these ion-photon interactions to herald entanglement between remote atomic memories. We offered simplified physical models to highlight tradeoffs between rate and fidelity. We find that the sub-$\unit{\milli\meter}$ cavities anticipated to enhance type-II entanglement rates may actually provide even larger rate improvements if refined to access the \ac{scr}. In particular, \ac{dit} protocols which verify atomic correlation by transmitting arriving photons promise high-bandwidth operation with near-unit fidelity. \Ac{cpf} protocols, which impose atom-controlled phase shifts on reflecting photons, can only achieve high fidelity if the incoming optical mode is perfectly matched to the cavity mode. If that condition is met, it can facilitate \ac{re} with even higher success probability, though a tighter interaction bandwidth forces a fidelity-rate tradeoff which can largely be ignored in the \ac{dit} scheme. The downside of either \ac{scr} protocol, relative to traditional type-II protocols, is a fundamental limit on protocol fidelity which does not afflict type-II schemes. Practically speaking, however, all protocols will be limited by non-protocol error sources, some outlined in \cref{sec:qbs}, which the models in \cref{sec:results} did not account for. These non-protocol error sources could easily dwarf the minute differences between protocol fidelities.

The potential gains in \ac{re} rate available from adopting \ac{scr}-based protocols are contingent on several experimental factors. In this paper, we highlighted mirror fabrication requirements. \Ac{dit} becomes practical with sub-$\unit{\milli\meter}$ cavities when non-transmissive losses can be reduced below $\mathcal{L}_B\lesssim 50\text{ppm}$, a readily achievable condition. To make \ac{cpf} viable, non-transmissive losses must be below $\mathcal{L}_B\lesssim 10\text{ppm}$ which is closer to the limit of current fabrication capabilities at optical wavelengths. Further refinements of laser-ablation based mirror fabrication or novel approaches e.g. \cite{jin2022micro} improve outcomes for all cavity-enhanced protocols, though the \ac{scr} protocols stand to benefit the most. 

This study assumed that coherent coupling rates of $g\sim 2\pi\times 65\unit{\mega\hertz}$ are achievable, based on fundamental limitations, but did not discuss the barriers to realizing this in practice. Chiefly, the ion must be well localized so that it samples the cavity field at its strongest location. This will involve continued research into mitigating the charging of dielectric mirror surfaces close to the ion \cite{harlander2010trapped} and increased motional heating rates also caused by these mirrors \cite{teller2021heating}. Transitioning to cryogenic trapping environments may be key to tackling the latter challenge. Beyond that, the standard prescription to improve $g$ calls for reducing cavity $R$, embracing transitions with a larger $\lambda\cdot R_{br}$ product, or designing cavities which situate the ion at a diffraction-limited mode waist. This last pathway requires innovation in current cavity optomechanics to maintain $\sim${\AA} alignment stability in ultra-high vacuum or cryogenic environments. 

Regarding protocol design, future studies might explore experimental techniques to reshape and refine the extended spectra of emitted photons and utilize this freedom as another system parameter to optimize protocol rate and fidelity.

\appendix*

\section{Atomic Transitions for Simulations}\label{apx:ba_trans}
For the simulations performed in \cref{sec:results}, all atomic dipole information assumed qubit configurations realized in $^{133}\text{Ba}^+$ and $^{137}\text{Ba}^+$ isotopes. We selected Barium as a target species because the $\lambda=493\unit{\nano\meter}$ and $\lambda=455\unit{\nano\meter}$ transitions have longer wavelengths than most commonly trapped ion species, allowing for higher quality mirror coatings and decreased sensitivity to surface roughness. Furthermore, we chose species featuring a hyperfine splitting so that, after heralding, we can transition the atomic states with a more stable splitting for storage. 

\begin{figure*}
    \centering
    \includegraphics[width=\linewidth]{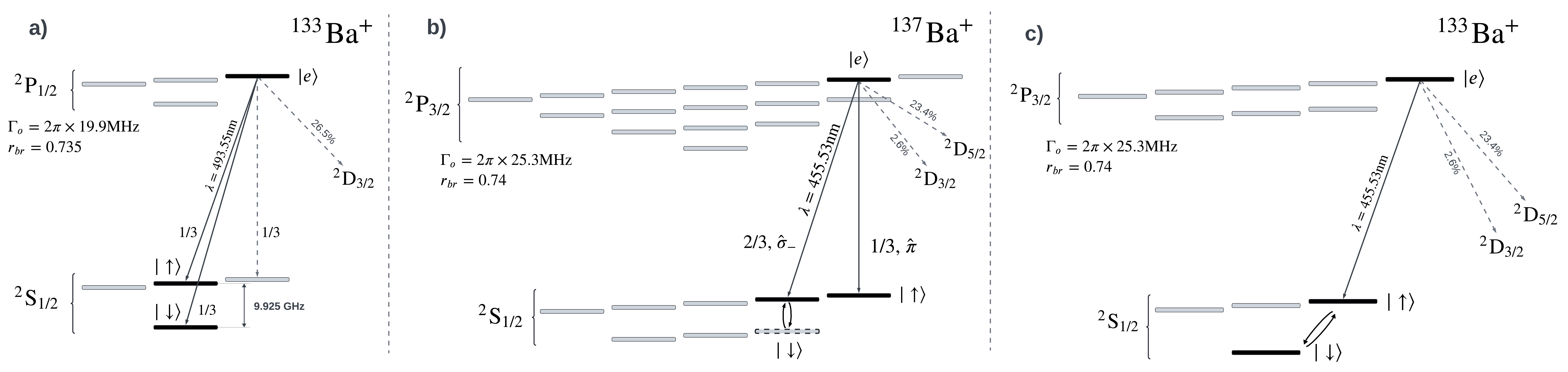}
    \caption{\label{fig:Excites}Ion-Photon entanglement schemes in various isotopes of Barium with Hyperfine splitting. Atomic states in these simplified level structures are labeled as $^{2j+\frac{1}{2}}\text{L}_{j}|F, m_F\rangle$ where $F=J\oplus I$ is the total angular momentum, a coupling between electronic angular momentum ($J$) and nuclear spin ($I$), and $m_F$ is the projection onto the quantization axis. (A) A double-resonance cavity with $\nu_{FSR}\approx\Delta_{HF}$ oriented along the quantization axis can simultaneously enhance $\hat{\sigma}_-$ polarized decays from the $^{2}\text{P}_{1/2}|1, 1\rangle$ to the $^{2}\text{S}_{1/2}|1, 0\rangle$ and $^{2}\text{S}_{1/2}|0, 0\rangle$, entangling $F$ with the photon frequency. The resulting frequency states are resolved by the hyperfine splitting $\Delta_{HF}\approx9.925\unit{\giga\hertz}$. (B) A cavity, resonant near $\lambda=455\unit{\nano\meter}$ and oriented orthogonal to the quantization axis enhances decays from the $^{2}\text{P}_{3/2}|3, 2\rangle$ to the $^{2}\text{S}_{1/2}|2, 1\rangle$ and $^{2}\text{S}_{1/2}|2, 2\rangle$ with $\hat{\sigma}_+$ and $\hat{\pi}$ polarized emissions respectively, entangling atom $m_F$ to the photon polarization. The cavity collects $\hat{\sigma}_-$ light half as effectively as $\hat{\pi}$ light, compensating for unequal free-space branching.  (C) A cavity, resonant near $\lambda=455\unit{\nano\meter}$ and aligned along the quantization axis, exclusively enhances the emission from the fully stretched $^{2}\text{P}_{3/2}|3, 2\rangle$ to the $^{2}\text{S}_{1/2}|1, 1\rangle$ in $^{133}\text{Ba}^+$. Atomic population may be dynamically removed from or to this coupled $^{2}\text{S}_{1/2}|1, 1\rangle$ between time-resolved excitation pulses, correlating the time of photon emission with the atomic state.}
\end{figure*}

\begin{table*}
    \caption{\label{tab:Scenario1} Transition Data. All $\lambda=493\unit{\nano\meter}$ $(455\unit{\nano\meter})$ transitions feature $\Gamma/2\pi = 19.9\unit{\mega\hertz}$ $(25.3\unit{\mega\hertz})$ full-width lines with $R_{br} = 0.735$ $(0.74)$. We also provide the effective transition dipole $\mu_{eff} = \sqrt{\sum_i\vec{\mu}_i\cdot\hat{\epsilon}_i}$, where $i$ indexes the ground states.  ($\dagger$) Non-hyperfine states are labeled in $|J, m_j\rangle$. ($\ddagger$) Alternative (and also more traditional) polarization collection scheme in a hyperfine qubit which represents only a 0.36\% decrease in interaction strength compared to the more atypical scheme we used.}
    \begin{ruledtabular}
    \begin{tabular}{cccccccccc} \hline
        \textrm{Type} &
        \textrm{Isotope} & 
        $|\text{e}\rangle$ & 
        $|\text{g}_1\rangle$ & 
        $|\text{g}_2\rangle$ & 
        Axis & 
        $\lambda$ $(\unit{\nano\meter})$ &
        $\vec{\mu}_i\cdot\hat{\epsilon}_i$ $(e a_o)$ &
        $\mu_{eff}$ $(e a_o)$\\
        \colrule
            Polarization & $^{137}\text{Ba}^+$ & $^2\text{P}_{3/2}|3,2\rangle$ & $^2\text{S}_{1/2}|2, 2\rangle$ & $^2\text{S}_{1/2}|2, 1\rangle$ & $\hat{x}$ & 455 & 1.353 & 1.913\\ 
            
            Frequency & $^{133}\text{Ba}^+$ & $^2\text{P}_{1/2}|1,1\rangle$ & $^2\text{S}_{1/2}|1, 0\rangle$ & $^2\text{S}_{1/2}|0, 0\rangle$ & $\hat{z}$ & 493 & 1.348 & 1.907\\ 
            
            Time-Bin & $^{133}\text{Ba}^+$ & $^2\text{P}_{3/2}|2,2\rangle$ & $^2\text{S}_{1/2}|1, 1\rangle$ & N/A & $\hat{z}$ & 455 & 2.343 & 2.343\\ \hline
            
            $\dagger$ Pol. (Trad) & $^{138}\text{Ba}^+$ & $^2\text{P}_{1/2}|\frac{1}{2},\frac{1}{2}\rangle$ & $^2\text{S}_{1/2}|\frac{1}{2}, \frac{1}{2}\rangle$ & $^2\text{S}_{1/2}|\frac{1}{2}, -\frac{1}{2}\rangle$ & $\hat{z}$ & 493 & 1.348 & 1.907\\ 
            
            $\dagger$ Time (Trad) & $^{138}\text{Ba}^+$ & $^2\text{P}_{1/2}|\frac{1}{2},\frac{1}{2}\rangle$ & $^2\text{S}_{1/2}|\frac{1}{2}, -\frac{1}{2}\rangle$ & N/A & $\hat{x}$ & 493 & 1.907 & 1.907\\ 
            
            $\ddagger$ Pol. (Alt) & $^{133}\text{Ba}^+$ & $^2\text{P}_{1/2}|1,0\rangle$ & $^2\text{S}_{1/2}|1,-1\rangle$ & $^2\text{S}_{1/2}|1, 1\rangle$ & $\hat{z}$ & 493 & 1.348 & 1.907\\ 
    \end{tabular}
    \end{ruledtabular}
\end{table*}

\begin{acknowledgments}
We thank Geert Vrijsen and Gyeonghun Kim for insightful discussions ans comments. This material is based upon work supported by the U.S. Department of Energy, Office of Science, National Quantum Information Science Research Centers, Quantum Systems Accelerator.
\end{acknowledgments}

\bibliography{ref}


\end{document}